\documentclass[12pt]{article}
\usepackage{titlesec}
\titleformat{\section}{\normalfont\Large\bfseries}{\thesection}{0.3em}{}
\titleformat{\subsection}{\normalfont\large\bfseries}{\thesubsection}{0.3em}{}
\titleformat{\subsubsection}{\normalfont\normalsize\bfseries}{\thesubsubsection}{0.3em}{}
\usepackage[nosort,noadjust]{cite}
\usepackage{amsmath,amsthm,amsfonts,amssymb,amscd,mathrsfs,slashed,graphicx,braket}
\usepackage{nicefrac,xfrac}
\usepackage{hyperref}
\usepackage[dvipsnames]{xcolor}

\newtheorem{lemma}{Lemma}
\usepackage{nicefrac,xfrac}
\hypersetup{
	colorlinks=true,
	citecolor=blue,
	linkcolor=blue,
	urlcolor=blue
}
\usepackage{cleveref}
\usepackage{tikz}
\usepackage{tikz-cd}
\usetikzlibrary{decorations.markings}
\usetikzlibrary{arrows.meta}
\tikzset{snake it/.style={decorate, decoration=snake}}
\usetikzlibrary{calc,decorations.markings}
\usetikzlibrary{decorations.pathmorphing}
\usetikzlibrary{decorations.pathreplacing}
\usetikzlibrary{shapes,backgrounds}
\usetikzlibrary{fadings}
\usetikzlibrary{cd}
\usetikzlibrary{hobby}
\usetikzlibrary{decorations.pathreplacing}
\usetikzlibrary{knots}
\usetikzlibrary{3d}
\tikzfading
[
name=fade out,
inner color=transparent!0,
outer color=transparent!100
]
\tikzset{
	partial ellipse/.style args={#1:#2:#3}{
		insert path={+ (#1:#3) arc (#1:#2:#3)}
	}
}
\newlength{\xtrawidth}
\newlength{\xtraheight}
\numberwithin{equation}{section}
\numberwithin{table}{section}
\numberwithin{figure}{section}
\newcommand{\RNum}[1]{\uppercase\expandafter{\romannumeral #1\relax}}
 \newcommand{\lr}[1]{\left( #1 \right)}
\newcommand{\thetas}[2]{\theta
	\begin{bmatrix}
		#1\\
     	#2
\end{bmatrix}}

\def\G2{\operatorname{G}_2}
\def\dd{{\rm d}}

\def\CFT#1{\operatorname{CFT}_{#1}}
\newcommand{\be}{\begin{equation}}
\newcommand{\ee}{\end{equation}}

\newcommand{\abs}[1]{\left|#1\right|}
\newcommand{\wt}[1]{\widetilde{#1}}
\newcommand{\mf}[1]{\mathfrak{#1}}

\newcommand{\nn}{\nonumber}

\begin{document}
\pagenumbering{Alph}
\begin{titlepage}
\vspace*{-2.5cm}
\begin{center}
\hfill BONN--TH--2024--07\\
\hfill MITP/24-034
\vskip 0.6in
{\LARGE\bf{Ensemble Averages of}}\\[2ex]
{\LARGE
$\boldsymbol{\mathbb{Z}_2}$~\bf{Orbifold Classes of Narain CFTs}}
\vskip 0.25in
{\bf
Stefan F\"orste$^{1}$,
Hans Jockers$^{2}$,
Joshua Kames-King$^{3,4}$,\\[0.5ex]
Alexandros Kanargias$^{2}$,
Ida G.~Zadeh$^{2}$}
\vskip 0.25in
{\it $^{1}\,$Bethe Center for Theoretical Physics, Physikalisches Institut der Universit\"at Bonn,\\
Nussallee 12, D-53115 Bonn, Germany}
\vskip 0.1in
{\it $^{2}\,$PRISMA+ Cluster of Excellence \& Mainz Institute for Theoretical Physics\\
Johannes Guttenberg-Universit\"at\\
Staudinger Weg 7, 55128 Mainz, Germany}
\vskip 0.1in
{\it $^{3}\,$Laboratory for Theoretical Fundamental Physics, Institute of Physics,\\
\'Ecole Polytechnique F\'ed\'erale de Lausanne, Switzerland}
\vskip 0.1in
{\it $^{4}\,$Fields and Strings Laboratory, Institute of Physics,\\
\'Ecole Polytechnique F\'ed\'erale de Lausanne, Switzerland}\\

\vskip 0.15in
{\tt forste@th.physik.uni-bonn.de}\\
{\tt jockers@uni-mainz.de}\\
{\tt jvakk@yahoo.com}\\
{\tt kanargias@uni-mainz.de}\\
{\tt zadeh@uni-mainz.de}
\end{center}
\vskip 0.15in
\begin{center} {\bf Abstract} \end{center}

In this work we study families of $\mathbb{Z}_2$~orbifolds of toroidal conformal field theories based on both factorizable and non-factorizable target space tori. For these classes of theories, we analyze their moduli spaces, and compute their partition functions. Building on previous work, we express the calculated partition functions in terms of suitable Siegel--Narain theta functions that allow us to determine their ensemble averages. We express the derived averaged partition functions of the studied families of conformal field theories in a manifest modular invariant finite sum of products of real analytic Eisenstein series. We speculate on a tentative holographic three-dimensional dual bulk interpretations for the considered $\mathbb{Z}_2$~orbifold classes of ensembles of conformal field theories.

\vfill
\noindent March 2024
\end{titlepage}
\pagenumbering{gobble}
\tableofcontents

\newpage
\pagenumbering{arabic}
\section{Introduction}
The AdS/CFT correspondence in its original form posits a duality between type~IIB string theory on $\text{AdS}_5 \times \text{S}^5$ and four-dimensional $\mathcal{N}=4$ SYM \cite{Maldacena:1997re}. Recent results however seem to indicate that low-dimensional theories of gravity are dual to some form of ensemble-average of quantum-mechanical theories rather than one individual theory. The most studied example to date is the duality between Jackiw--Teitelboim gravity in two dimensions and a specific double-scaled matrix integral \cite{Jackiw:1984je,Teitelboim:1983ux,Saad:2019lba}. There is some evidence that a similar relationship might hold between three-dimensional Einstein gravity and an appropriately defined ensemble of two-dimensional conformal field theories (CFTs).\footnote{For a recent discussion of the role of ensemble averages in the context of the original AdS/CFT correspondence between $\mathcal{N}=4$ SYM and the gravitational bulk theory on $\text{AdS}_5$, see ref.~\cite{Collier:2022emf}. Further considerations of ensemble averages of quantum field theories in various dimensions arising from string theory appear in refs.~\cite{Heckman:2021vzx, Baume:2023kkf}. In ref.~\cite{Balasubramanian:2020lux} phenomenological implications of ensemble averages of quantum field theories are considered.} See refs.~\cite{Belin:2020hea,Cotler:2020ugk,Chandra:2022bqq,DiUbaldo:2023qli,Belin:2023efa} for progress in that direction. A telltale sign of such averaged dualities is the presence of Euclidean wormhole geometries in the gravitational path integral. These destroy factorisation of the dual partition function~\cite{Maldacena:2004rf}, suggesting an ensemble interpretation. This is in stark contrast to examples derived from string theory such as the aforementioned original duality. It is an open question how the bottom-up and top-down approaches are reconciled. See refs.~\cite{Eberhardt:2021jvj,Benjamin:2021wzr,Collier:2022emf,Kames-King:2023fpa,Ashwinkumar:2023jtz,Schlenker:2022dyo}.

In refs~\cite{Maloney:2020nni,Afkhami-Jeddi:2020ezh} a setting is considered, in which a precise definition can be given to an ensemble of two-dimensional conformal field theories. Namely, in those works the ensemble arises from $D$ free massless two-dimensional bosons parametrizing a target space torus $T^D$, which is referred to as the ensemble of Narain conformal field theories. The moduli space $\mathcal{M}_{T^D}$ of this Narain family carries a natural measure in the form of the Zamolodchikov metric \cite{Zamolodchikov:1986gt}. To determine ensemble-averaged quantities in the Narain ensemble with respect to this measure --- such as the ensemble-averaged partition function \cite{Maloney:2020nni,Afkhami-Jeddi:2020ezh} --- one uses the Siegel--Weil theorem \cite{Siegel,siegel1955lectures,Weil1,Weil2}, which plays a central role in our work as well.
Due to the $U(1)$ symmetry currents in the family of toroidal conformal field theories, the three-dimensional dual holographic bulk theory is expected to be enhanced by a gauge symmetry. Indeed, in refs.~\cite{Maloney:2020nni,Afkhami-Jeddi:2020ezh} a holographic interpretation for the Narain ensemble is proposed in terms of a $U(1)^{D}\times U(1)^{D}$ Chern--Simons theory summed over three-manifolds. While the Chern--Simons theory may a priori be defined on any three-manifold, the specific choice of what to include in such a sum is dictated by the Siegel--Weil formula. Note that it is the described symmetry enhancement, which offers a method to calculate ensemble averaged quantities in the dual bulk theory. Thus, the given holographic formulation in the form of the $U(1)^{D}\times U(1)^{D}$ Chern--Simons theory is simpler than Einstein gravity, such that the described holographic correspondence becomes tractable (at least at a computational level).
Various works extend the holographic correspondence between the Narain ensemble of conformal field theories and the bulk $U(1)^{D}\times U(1)^{D}$ Chern--Simons theories. For additional results on the original Narain duality see refs.~\cite{Maloney:2020nni,Afkhami-Jeddi:2020ezh}, and for various further extensions, see refs.~\cite{Datta:2021ftn,Collier:2021rsn,Benjamin:2021wzr,Ashwinkumar:2021kav,Dong:2021,Perez:2020klz,Dymarsky:2020qom,Dymarsky:2020pzc,Dymarsky:2021xfc,Meruliya:2021utr,Meruliya:2021lul,Kawabata:2023iss,Aharony:2023zit,Ashwinkumar:2023ctt,Barbar:2023ncl,Raeymaekers:2023ras,Kames-King:2023fpa}. Particularly relevant to our work, in ref.~\cite{Benjamin:2021wzr} a duality is established between specific families of $\mathbb{Z}_N$~orbifolds of Narain conformal field theories and $U(1)^{D}\times U(1)^{D}$ Chern--Simons theories plus an additional discrete $\mathbb{Z}_N$ gauge group factor. On the conformal field theory side, the twisted sectors of the $\mathbb{Z}_N$~orbifold enter in the ensemble average of the partition function, whereas on the dual side of the Chern--Simons theory vortex configurations --- originating from the discrete $\mathbb{Z}_N$ gauge theory factor --- account for these additional contributions. While in ref.~\cite{Benjamin:2021wzr} the studied $\mathbb{Z}_N$ orbifolds act uniformly on all torus directions, we focus in this work on $\mathbb{Z}_2$~orbifolds of toroidal conformal field theories whose geometric action on their target space tori is more general. These more general classes of $\mathbb{Z}_2$ orbifold theories fall also into the ensembles considered in the interesting work~\cite{Ashwinkumar:2023ctt}, which discusses ensembles of theories resulting from Narain lattices with arbitrary signatures and orbifolds thereof from a more general but less geometric point of view. For an interesting relationship between orbifolds of Narain conformal field theories and the constructions of Narain conformal field theories from quantum codes see the recent developments in refs.~\cite{Dymarsky:2020qom,Dymarsky:2020pzc,Dymarsky:2021xfc,Kawabata:2023iss}.

We distinguish between two important families of toroidal $\mathbb{Z}_2$~orbifold conformal field theories that are based on factorizable or non-factorizable target space tori.\footnote{Topologically, any torus of arbitrary dimension factorizes into tori of smaller dimension, e.g., $T^D\simeq_{\text{top.}}T^\ell\times T^m$ for $D=\ell+m$. However, as a Riemannian manifold it only factorizes in this way when the metric is block diagonal.}
These two classes of conformal field theories come with two topologically distinct $\mathbb{Z}_2$~orbifold group actions. We find that the former class of theories is the product family of the ensemble of Narain conformal field theories studied in refs.~\cite{Maloney:2020nni,Afkhami-Jeddi:2020ezh} and the ensemble of $\mathbb{Z}_2$~orbifolds of the Narain conformal field theories analyzed in ref.~\cite{Benjamin:2021wzr}. The latter class of theories enlarges the ensemble of $S_2$~symmetric toroidal orbifold conformal field theories analyzed in ref.~\cite{Kames-King:2023fpa}. Namely, the moduli space of the ensemble of $S_2$~symmetric orbifold conformal field theories forms a half-dimensional subspace in the moduli space of non-factorizable $\mathbb{Z}_2$~orbifold toroidal conformal field theories examined here.

In order to calculate the ensemble averages $\left\langle Z(\tau) \right\rangle$ of the conformal field theory partition function in terms of the worldsheet modular parameter $\tau$, it is necessary to first determine the moduli space for the family of conformal field theories together with a measure. For the original ensemble of Narain conformal field theories of the target space torus $T^D$ of refs.~\cite{Maloney:2020nni,Afkhami-Jeddi:2020ezh}, the moduli space $\mathcal{M}_{T^D}$ is well-known and realized by the homogeneous space 
\begin{equation}
\mathcal{M}_{T^D} \simeq \left.\raisebox{-0.75ex}{$O(D,D,\mathbb{Z})$} \middle\backslash {\raisebox{0.75ex}{$O(D,D,\mathbb{R})$}} \middle\slash 
     \raisebox{-0.75ex}{$O(D,\mathbb{R}) \times O(D,\mathbb{R})$}\right. \ .
\end{equation}
A measure $\dd\mu(\mathfrak{m})$ is obtained from the Zamolodchikov metric of a member $\mathfrak{m} \in \mathcal{M}_{T^D}$ in the ensemble of Narain theories. Observing that the moduli dependence of the partition function $Z_{T^D}(\tau,\mathfrak{m})$ is entirely captured by its proportional Siegel--Narain Theta function $\Theta(\tau,\mathfrak{m})$, calculating the ensemble average of the partition function amounts to averaging the Siegel--Narain Theta function $\Theta(\tau,\mathfrak{m})$ \cite{Maloney:2020nni,Afkhami-Jeddi:2020ezh}. This average is determined by the Siegel--Weil formula \cite{Siegel,siegel1955lectures,Weil1,Weil2}, which yields 
\begin{equation}
    \int_{\mathcal{M}_{T^D}} \dd \mu(\mathfrak{m}) \, \Theta(\tau,\mathfrak{m})\, 
    = \frac{E_{\nicefrac{D}{2}}(\tau)}{\left(\operatorname{Im} \tau\right)^{\nicefrac{D}{2}}}
    \quad \text{for} \quad D\ge 3 \ .
\end{equation}
Here $E_{\nicefrac{D}{2}}(\tau)$ is the real analytic Eisenstein function, which is a modular function with respect to the modular parameter $\tau$ of the modular group $\operatorname{PSL}(2,\mathbb{Z})$. 

To describe the ensemble of $\mathbb{Z}_2$~orbifolded Narain conformal field theories, we first determine the moduli spaces $\mathcal{M}_{T^D/\mathbb{Z}_2}$. While the $\mathbb{Z}_2$~orbifold of the Narain theories considered in ref.~\cite{Benjamin:2021wzr} are defined for any modulus $\mathfrak{m}$ of $\mathcal{M}_{T^D}$, the generalized class of $\mathbb{Z}_2$~orbifolds of this work cannot be realized for generic moduli $\mathfrak{m}$ of $\mathcal{M}_{T^D}$. That is to say, the relevant moduli spaces $\mathcal{M}_{T^D/\mathbb{Z}_2}$ are those subspaces of $\mathcal{M}_{T^D}$, for which the $\mathbb{Z}_2$~orbifold symmetry exists in the unorbifolded Narain theory. We find that for both the factorizable and the non-factorizable $\mathbb{Z}_2$~orbifold classes, the moduli space $\mathcal{M}_{T^D/\mathbb{Z}_2}$ becomes a product $\mathcal{M}_{T^\ell} \times \mathcal{M}_{T^m}$ with $D=\ell+m$ for suitable $m$ and $\ell$, on which the Zamolodchikov measure $\dd \mu$ factorizes accordingly. 

Inspired by the interesting work~\cite{Dong:2021}, we express the partition functions of $\mathbb{Z}_2$~orbifold toroidal conformal field theories in terms of sums of products of a broader class of Siegel--Narain Theta function~$\Theta_{H}(a,b,\tau)$ to be defined in the main text. This class of Siegel--Narain Theta functions is indeed required to get a handle on the ensemble averages in particular for the class of non-factorizable $\mathbb{Z}_2$~orbifolds. Building on the results of ref.~\cite{Dong:2021} (see also ref.~\cite{Ashwinkumar:2023ctt}), the products of these Siegel--Narain Theta functions respect the product structure of moduli spaces, such that the ensemble averages of the derived partition functions are again calculable with the Siegel--Weil formula.

In the original Narain correspondence the appearance of the real analytic Eisenstein series~$E_{\nicefrac{D}{2}}(\tau)$ in the ensemble average of the partition function leads to the proposal of a three-dimensional holographic dual bulk interpretation of the form \cite{Maloney:2020nni,Afkhami-Jeddi:2020ezh}
\begin{equation} \label{eq:Sum3d}
  \left\langle Z_{T^D}(\tau) \right\rangle = \sum_{\text{3-manifolds}} \int \mathcal{D}[A,{\wt A}] \, e^{-S_{\text{bulk}}[A,{\wt A}]} \ .
\end{equation}
The bulk action $S_{\text{bulk}}[A,{\wt A}]$ is realized by an Abelian $U(1)^D \times U(1)^D$ Chern--Simons theory, where the gauge connections $A_a$ and ${\wt A}_a$, $a=1,\ldots,D$, arise from the global $U(1)$~currents of the toroidal conformal field theory at the asymptotic boundary.\footnote{Note that it is not quite clear in how far the theory can be defined non-perturbatively for various reasons \cite{Maloney:2020nni}. It might be more appropriate to consider the bulk to be an effective description with a UV-completion given in terms of a specific member of the ensemble \cite{Dong:2021}.}
Upon expressing the real analytic Eisenstein series~$E_{\nicefrac{D}{2}}(\tau)$ as a suitable sum over $\operatorname{PSL}(2,\mathbb{Z})$-modular orbits, the real analytic Eisenstein series~$E_{\nicefrac{D}{2}}(\tau)$ in the ensemble average suggests a sum over three manifolds in the three-dimensional formulation~\eqref{eq:Sum3d}~\cite{Maloney:2020nni,Afkhami-Jeddi:2020ezh}. Namely, real analytic Eisenstein series $E_{\nicefrac{D}{2}}(\tau)$ appear as a sum over inequivalent hyperbolic genus one handle-bodies \cite{Maldacena:1998bw,Maloney:2007ud,Manschot:2007ha,Keller:2014xba}.
As the orbifold constructions amount to the additional gauging of a discrete symmetry present in the original theory, it suggests that the bulk Chern--Simons theory has additional discrete gauge group factors \cite{Benjamin:2021wzr,Kames-King:2023fpa}. As explained in ref.~\cite{Benjamin:2021wzr} twist operators of the orbifolded conformal field theory are implemented as vortices in the three-dimensional bulk theory, which are line operators implementing non-trivial boundary conditions around contractible cycles in the bulk manifold \cite{PhysRevLett.62.1221,PRESKILL199050}. 

For the ensemble averages of the partition functions of the families of $\mathbb{Z}_2$~orbifold toroidal conformal field theories studied in ref.~\cite{Benjamin:2021wzr}, including the vortex sectors associated to the additional discrete $\mathbb{Z}_2$~gauge group generalizes the Narain holographic correspondence~\eqref{eq:Sum3d}. However, the fact that the ensembles of $\mathbb{Z}_2$~orbifold conformal field theories studied in this work constrain the moduli space $\mathcal{M}_{T^D}$ to a product sub-moduli space $\mathcal{M}_{T^\ell} \times \mathcal{M}_{T^{m\vphantom{\ell}}}$ renders a three-dimensional dual holographic bulk interpretation more challenging, see also ref.~\cite{Benjamin:2021ygh}. At the technical level the three-dimensional bulk theory must now reproduce (sums of) products of real analytic Eisenstein series that appear in the expressions of the ensemble average of the conformal field theory partition functions $\left\langle Z_{T^D/\mathbb{Z}_2}(\tau) \right\rangle$. Under the assumption that there is a holographic dual description we formulate possible implications for the dual bulk theory that are a consequence of the observed product structure of the moduli spaces.

\subsection*{Outline of Results}
Let us now outline the rest of the paper and reference the main results of the following sections:

In section~\ref{sec:ObCFTT2} we describe the general logic in constructing $\mathbb{Z}_2$ orbifold actions by considering $\mathbb{Z}_2$~orbifold conformal field theories for two-dimensional target space tori~$T^2$. This basic setup already allows us to introduce the notion of factorizable and non-factorizable $\mathbb{Z}_2$~orbifolds that are the key player of this work. For these two classes of orbifold theories based on target space torus $T^2$, we determine their partition functions and study their ensembles. Due to the low dimensionality of the target space torus $T^2$, the ensemble average of their partition functions strictly speaking diverges. However, similarly as in ref.~\cite{Maloney:2020nni}, upon considering regularized ensemble averages, we are able to exhibit already in this simple setup the general structure of ensemble averages for factorizable and non-factorizable $\mathbb{Z}_2$~orbifold theories, which is a helpful guidline in the following sections. 

In section~\ref{sec:ObCFTTD} we study both the factorizable and the non-factorizable families of $\mathbb{Z}_2$~orbifold conformal field theories with higher dimensional toroidal target spaces, for which the volumes of their moduli spaces and the ensemble averages of their partition functions are finite. We find that the moduli spaces for both the factorizable and the non-factorizable classes exhibit a product structure. We establish that for the class of factorizable $\mathbb{Z}_2$ orbifolds the resulting ensemble average of the partition function factorizes into two contributions. These two factors correspond to the ensemble averages of the partition functions of the Narain conformal field theories studied in ref.~\cite{Maloney:2020nni,Afkhami-Jeddi:2020ezh} and of the $\mathbb{Z}_2$~orbifold conformal field theories studied in ref.~\cite{Benjamin:2021wzr}. The partition function for the non-factorizable $\mathbb{Z}_2$~orbifold toroidal conformal field theories generalizes the result for the $S_2 \simeq \mathbb{Z}_2$~symmetric orbifolds studied in ref.~\cite{Kames-King:2023fpa}, in the sense that the moduli space of the non-factorizable $\mathbb{Z}_2$~orbifold family embeds into the moduli space of the family of the $S_2$~symmetric orbifold conformal field theory as a subslice. Following ref.~\cite{Dong:2021}, we express the partition function of the non-factorizable $\mathbb{Z}_2$~orbifold theory in terms of suitable Siegel--Narain Eisenstein series, which allow us to calculate their ensemble averages with the Siegel--Weil formula. In this way, we arrive at manifest modular invariant ensemble averages that are given as a sum of products of real analytic Eisenstein series.

In section \ref{sec:Bulk} we make tentative comments about a possible holographic bulk interpretation of the calculated ensemble averages of the partition functions derived in the previous section. In particular, we discuss the following two scenarios: On the one hand, the holographic dual interpretation can be disguised by considering a less suitable ensemble of conformal field theories in the first place. This scenario is for instance suggested if the moduli space of the considered ensemble embeds into a larger moduli space. The hallmark of such a scenario is the presence of additional exactly marginal operators in the ensemble of conformal field theories that do not parametrize directions tangent to the considered moduli space. On the other hand, the product structure of the moduli spaces of the conformal field theories in this work could just be a feature that the three-dimensional bulk theory needs to reproduce. The consequence of this scenario is that the holographic dual three-dimensional bulk geometries arise from a pair of three-spaces that are glued together at a common asymptotic toroidal boundary.

Finally, in section~\ref{sec:con} we present our conclusions and discuss open questions and further research directions. Some technical details of our computations are relegated to three appendices.

\section{Orbifold CFTs from Two-Dimensional Tori} \label{sec:ObCFTT2} 
In this section we systematically study two-dimensional conformal field theories arising from $\mathbb{Z}_2$ orbifolds of two-dimensional tori $T^2$. Let $T^2$ be the two-dimensional torus
\begin{equation} \label{eq:2torus}
   T^2 \simeq \mathbb{C} / ( \mathbb{Z} + u \mathbb{Z} ) \ ,
\end{equation}
in terms of the complex structure parameter $u$ in the upper half-plane $\mathcal{H}$, i.e., $\operatorname{Im}u >0$. The volume modulus~$k$ together with the antisymmetric tensor field~$B$ forms the complexified K{\"a}hler modulus
$t = \frac{1}{\alpha'}\lr{B + i k}:=b+i\kappa$. Here the dimensionless parameters $b$ and $\kappa$ parametrize the two-form background $B$-field
and the positive K\"ahler two-form, respectively, such that the complexified K\"ahler parameter $t$ also takes values in the upper half-plane $\mathcal{H}$. For further details on these conformal field theories and their moduli see, e.g., ref.~\cite{Blumenhagen:2013fgp}.

The toroidal orbifold $\mathbb{Z}_2$ action is generated by an involution $\iota_{\mathbb{Z}_2}: T^2 \to T^2$, i.e., $\iota_{\mathbb{Z}_2}^2 = \mathbf{1}$, which induces a $\mathbb{Z}_2$ action on the two-form $B$-field via pullback $\iota^*_{\mathbb{Z}_2}$. We do not consider fixed-point free involutions $\iota_{\mathbb{Z}_2}$ --- referred to as shift orbifolds in ref.~\cite{Wendland:2000ye} --- because the associated free $\mathbb{Z}_2$ action simply yields another toroidal CFT of $T^2$ (with adjusted background parameters $u$ and $t$).

Involutions $\iota_{\mathbb{Z}_2}$ with non-trivial fixed points arise from reflections about points and lines, which must respectively be symmetry points and symmetry axes of the corresponding torus $T^2$. As the lattice point $z\in \mathbb{Z} + u \mathbb{Z}$ is always a symmetry point of reflection with respect to the origin in $T^2$, the associated toroidal $\mathbb{Z}_2$~orbifold is well-defined for any choice of moduli $u$ and $t$. Such orbifold conformal field theories and their ensembles are studied in detail in ref.~\cite{Benjamin:2021wzr}, and are therefore not further discussed here. 

\begin{figure}
		\begin{center}
			\begin{tikzpicture}[scale=3]
				
				\draw[->] (-0.2,0) -- (1.8,0) ;
				\draw[->] (0,-0.2) -- (0,1.5) ;
				\draw (0.2,0) arc (0:90:0.2);
                \draw [fill] (0.075,0.075) circle [radius=0.3pt];
				\node[above right] at (0.25,0.01) {$\phi$};
				\draw[->, thick] (0,0) -- (1,0) ;
				\draw[->, thick] (0,0) -- (0,1.4) ;
				\draw[very thick]  ({cos(90)},1.4) -- ({1+cos(90)},1.4) ;
				\draw[very thick]  (1,0) -- ({1+cos(90)},1.4) ;
				\draw[ultra thick, magenta] (0.5,0) -- (0.5,1.4) ;
				\draw[ultra thick, magenta] (0,0) -- (0,1.38) ;
                \draw[ultra thick, dashed, Cyan] (0,.7) -- (1,0.7) ;
				\draw[ultra thick, dashed, Cyan] (0,0) -- (1,0) ;
				\node[below] at (1.1,0) {$1$};
				\node[below] at (0.5,0) {$\frac12$};
				\node[left] at (0,1) {$u$};
				\node[right] at (1.4,1.4) {\framebox{$\mathbb C$}};
			\begin{scope}[scale=1,xshift=2.5cm]
				\draw[->] (-0.2,0) -- (1.8,0) ;
				\draw[->] (0,-0.2) -- (0,1.5) ;
				\draw (0.2,0) arc (0:35:0.2);
				\node[below right] at (0.45,0.2) {$\phi$};
				\draw[->, ultra thick] (0,0) -- (1,0) ;
				\draw[->, ultra thick] (0,0) -- ({cos(35)},{sin(35)}) ;
				\draw[very thick]  ({cos(35)},{sin(35)}) -- ({1+cos(35)},{sin(35)}) ;
				\draw[very thick]  (1,0) -- ({1+cos(35)},{sin(35)}) ;
				\draw[ultra thick, orange] (0,0) -- ({1+cos(35)},{sin(35)}) ;
				\draw[ultra thick, dashed, green] (1,0) -- ({cos(35)},{sin(35)}) ;
				\node[below] at (1.1,0) {$1$};
				\node[above left] at ({cos(35)},{sin(35)}) {$u$};
				\node[right] at (1.4,1.4) {\framebox{$\mathbb C$}};
				\end{scope}
			\end{tikzpicture}
		\end{center}
		\caption{Factorizable and non-factorizable lattices. On the left, the two fixed point loci of the reflection along each lattice generator are depicted with solid magenta and dashed cyan lines. On the right, the symmetry axes along the diagonal lines are depicted with solid orange and dashed green lines.}\label{fig: fact and non fact}
\end{figure}
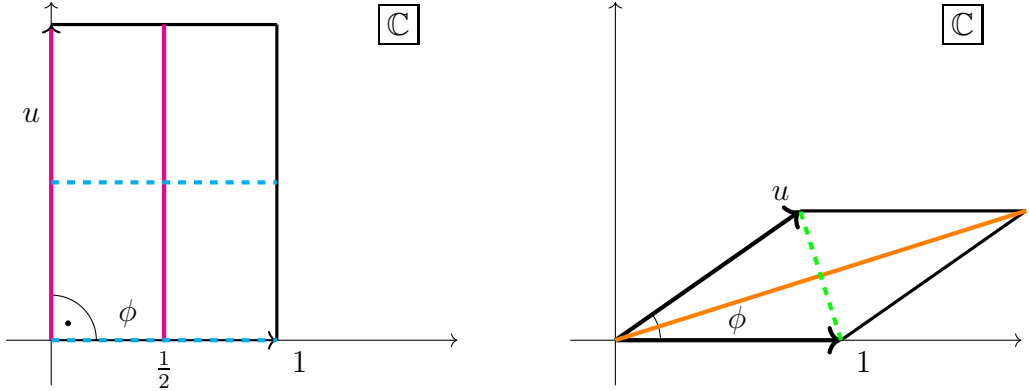

Reflections along a line of $T^2$ are what we are interested in. These are only possible if the line is a symmetry axis of the torus $T^2$, which imposes constraints on the moduli $u$ and $t$. There are two types of symmetry axes of two-dimensional tori. Firstly, we can reflect along an axis associated to a lattice generator of $\mathbb{Z} + u \mathbb{Z}$, which amounts to a reflection along the bounding edge of a primitive cell of the torus lattice. For this axis to be a symmetry, the primitive cell must be a rectangle (see the left panel of Fig. \ref{fig: fact and non fact}). One may reflect with respect to either of the two axes, and obtain the two fixed point loci associated with each reflection.

Secondly, we may reflect along an axis associated to a lattice vector that is the sum of, or the difference between, the two distinct generators of $\mathbb{Z} + u \mathbb{Z}$ (see the right panel of Fig. \ref{fig: fact and non fact}). This realizes reflections along the two diagonals of the primitive cell spanned by these two generators. These axes are only a symmetry of the torus if the primitive cell is a rhombus.

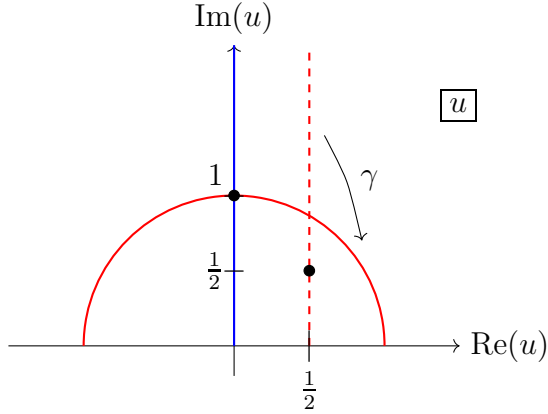
\begin{figure}
\begin{center}
		\begin{tikzpicture}[scale=2]
			\draw[->] (-1.5,0) -- (1.5,0) node[right] {$\text{Re}(u)$} ;
			\draw[->] (0,-0.2) -- (0,2.) node[above] {$\text{Im}(u)$};
			\node[right] at (1.3,1.6) {\framebox{$u$}};
			\draw[thick,red] (0,0) [partial ellipse = 0:180:1 ];
			\draw[thick,red,dashed] (0.5,0) -- (0.5, 2);
			\draw[thick,blue] (0,0) -- (0, 2);
			\draw[->] plot [smooth] coordinates {(0.6,1.4) (0.75,1.09)(0.85,0.7)};
			\node at (0.9,1.1) {$\gamma$};
			\draw [black,fill] (0.5,0.5) circle [radius=1pt];
			\draw [black,fill] (0,1) circle [radius=1pt];
			\node at (0.5,0) {$\lvert$};
			\node[below] at (0.5,-0.1) {$\frac12$};
			\node at (0,0.998) {$-$};
			\node[above left] at (0,1) {$1$};
			\node at (0,0.5) {$-$};
			\node[left] at (0,0.5) {$\frac12$};
		\end{tikzpicture}
	\end{center}
	\caption{Complex structure plane. The blue line and the solid red arc correspond, respectively, to the factorizable and non-factorizable lattice. The dashed red line is mapped to the red arc by the modular transformation $\gamma=(u-1)/u$ (see footnote \ref{ft:shiftlattice}). The two black dots are mapped to each other under $\gamma$.}\label{fig: moduli space orbifold}
	\end{figure}

The former configurations with a rectangular primitive cell are known as factorizable tori, whereas the latter configurations are often referred to as non-factorizable tori. These two possibilities realize two distinct classes of configurations that are characterized by the complex structure modulus $u$, which for the factorizable tori is constrained to be purely imaginary, i.e., $\operatorname{Re}(u)=0$, and for the non-factorizable tori is constrained to be a phase, i.e., $| u | = 1$.\footnote{Actually, $\operatorname{Re}(\tilde u) = \frac12$ also realizes non-factorizable tori because the lattice generators $\tilde u$ and $\tilde u - 1$ form a rhombus in this case. These two sides of the rhombus are exchanged by a reflection along the imaginary axis. This alternative description is often used in the literature and it is related to $|u |=1$ via the M\"obius transformation $u = (\tilde u - 1)/{\tilde u}$ in complex structure moduli space\label{ft:shiftlattice} (see Fig. \ref{fig: moduli space orbifold}).} Note that the square torus with $u = i$ admits both a factorizable and a non-factorizable orbifold action.

The involution $\iota_{\mathbb{Z}_2}$ for the reflections of either factorizable tori or non-factorizable tori induces a $\mathbb{Z}_2$-action on the $B$-field (via the pull-back $\iota^*_{\mathbb Z_2}$), which maps $b = \operatorname{Re}(t)$ to $-b$. Due to the periodicity of the $B$-field, $b \sim b+1$, there are two possible invariant choices for the background value of $b$, namely $b=0$ or $b=\frac{1}2$. Hence, with $\operatorname{Re}(t)=0$ and $\operatorname{Re}(t)=\frac12$ there are two possible classes of background values for the complexified K\"ahler modulus $t$, which admit the discussed $\mathbb{Z}_2$ orbifold action. That there are two possibilities for the K\"ahler modulus is not a coincidence as mirror symmetry of $T^2$ maps the factorizable torus in complex structure moduli space to a configuration with vanishing $B$-field in the complexified K\"ahler moduli space, and the non-factorizable torus in complex structure to a half-integral $B$-field in K\"ahler moduli space, and vice versa. 

The family of conformal field theories of the two-dimensional tori~$T^2$ is described by the moduli space (see for instance ref.~\cite{Blumenhagen:2013fgp})
\begin{equation} \label{eq:ModT2}
    \mathcal{M}_{T^2}=\lr{\mathcal{H}/\text{PSL}\lr{2,\mathbb{Z}_2}\times \mathcal{H}/\operatorname{PSL}\lr{2,\mathbb{Z}_2}}/\lr{\mathbb{Z}_2\times \mathbb{Z}_2} \ .
\end{equation}
The moduli space $\mathcal{M}_{T^2/\mathbb{Z}_2}$ of the family of toroidal $\mathbb{Z}_2$~orbifold conformal field theories associated to the involution $\iota_{\mathbb{Z}_2}$ is the subspace of the moduli space $\mathcal{M}_{T^2}$, which parametrizes those two-dimensional tori~$T^2_{(u,t)}$ that admit the involution~$\iota_{\mathbb{Z}_2}$ as a $\mathbb{Z}_2$~symmetry, i.e.,
\begin{equation} \label{eq:InvModSpace}
   \mathcal{M}_{T^2/\mathbb{Z}_2} =
   \left\{ (u,t) \in \mathcal{M}_{T^2} \,\middle|\,
   \exists \ \iota_{\mathbb{Z}_2} \ \text{ on } T^2_{(u,t)} \right\} 
   \subset \mathcal{M}_{T^2} \ . 
\end{equation}
Here $T^2_{(u,t)}$ denotes for a given $(u,t)\in\mathcal{M}_{T^2}$ the two-dimensional torus with complex structure $u$ and K\"ahler structure $t$.

\subsection{Factorizable \texorpdfstring{$\boldsymbol{\mathbb{Z}_2}$}{Z2} Orbifold} \label{sec:factorb}

Let us first consider the class of factorizable $\mathbb{Z}_2$ orbifold theories with a vanishing background $B$-field. That is to say, the orbifold theory is given in terms of a two-dimensional torus $T^2$ of a rectangular primitive cell together with the involution~$\iota_{\mathbb{Z}_2}$ that arises from a reflection along the imaginary axis in the toroidal universal covering space $\mathbb{C}$.\footnote{Reflection along the real axis forms another involution of such a factorizable torus. However, the orbifold theories arising from either one of these involutions are equivalent because upon rotating and conformally rescaling the lattice $\mathbb Z+u\mathbb Z$ of the two-dimensional torus to the conformally equivalent lattice $\mathbb Z+u^{-1}\mathbb Z$ exchanges these two involutions.}
This family of torodial $\mathbb{Z}_2$~orbifold theories is then parametrized by an imaginary complex structure modulus and an imaginary complexified K\"ahler modulus, i.e.,
\begin{equation} \label{eq:modfact2torus}
  u = i c \ , \qquad t = i\kappa\ ,\qquad 1\le c<+\infty\ ,\qquad 1\le \kappa<+\infty\ ,
\end{equation}
with the real complex structure parameter $c$ and the real K\"ahler modulus $\kappa$. The moduli space $\mathcal{M}_{T^2/\mathbb{Z}_2}$ spelt out in eq.~\eqref{eq:InvModSpace} therefore becomes
\begin{equation}
  \mathcal{M}_{T^2/\mathbb{Z}_2} \simeq [1,+\infty)^2/\mathbb{Z}_2 \subset
  \mathcal{M}_{T^2} \ .
\end{equation}
Here, the remaining $\mathbb Z_2$ quotient corresponds to the exchange of $(c,\kappa)\leftrightarrow (\kappa,c)$.
The fixed-point locus of the involution $\iota_{\mathbb{Z}_2}$ has  two disconnected components, namely the two circles $\operatorname{Re}(z) = 0$ and $\operatorname{Re}(z) = \frac12$, where $z$ is the complex coordinate of the universal covering space $\mathbb{C}$ of the two-dimensional torus~\eqref{eq:2torus}. Both fixed circles realize the same generator in the homology group $H_1(T^2,\mathbb{Z})$.

The described $\mathbb{Z}_2$ orbifold therefore factors as
\begin{equation} \label{eq:fact2torus}
T^2/\mathbb{Z}_2 \simeq S^1/{\mathbb{Z}_2} \times S^1 \ , 
\end{equation}
where the first and second factors on the right hand side are parametrized in terms of the real and imaginary part of $z$, respectively. Furthermore, we denote by $R_1$ and $R_2$ the radii of these two respective circles, which in terms of the parameters~\eqref{eq:modfact2torus} are given by 
\begin{equation}
2\pi R_1 = \sqrt{\frac{\alpha'\kappa}{c}}  \ , \qquad  2\pi R_2 =  \sqrt{ \alpha'c \kappa} \ ,
\end{equation}
where we used the metric \begin{equation}
    g=\begin{pmatrix}
        \lr{2\pi R_1}^2 & 0\\
        0 & \lr{2\pi R_2}^2
    \end{pmatrix}\ .
\end{equation}

As a result, the partition function $Z_{T^2_\text{fac}/\mathbb{Z}_2}$ of the factorizable $\mathbb{Z}_2$ orbifold conformal field theory factorizes accordingly:
\begin{equation} \label{eq:Fact2PF}
   Z_{T^2_\text{fac}/\mathbb{Z}_2}(\tau; c,\kappa) =   Z_{S^1/{\mathbb{Z}_2}}(\tau; R_1(c,\kappa)) \, Z_{S^1}(\tau; R_2(c,\kappa)) \ ,
\end{equation}   
where  $Z_{S^1}$ and $Z_{S^1/\mathbb{Z}_2}$ are the partition functions of the conformal field theories arising from a free boson on the circle $S^1$ and on the circle orbifold $S^1/\mathbb{Z}_2$, respectively. These partition functions are well-known (see e.g., refs.~\cite{Polchinski:1998rq,Blumenhagen:2013fgp}) and are given by
\begin{equation} \label{eq:PFS1}
    Z_{S^1}(\tau; R) = \frac{1}{|\eta(\tau)|^2} \sum_{w, m\in \mathbb{Z}} 
    e^{-2 \pi i \tau_1 w m - \pi \tau_2 \left( \frac{\alpha'}{R^2} w^2 + \frac{R^2}{\alpha'} m^2 \right) }
    \ ,
\end{equation}
and
\begin{equation}
    Z_{S^1/{\mathbb{Z}_2}}(\tau; R)  = \frac12 Z_{S^1}(\tau, R) 
    + \left( \left|\frac{\eta(\tau)}{\theta_2(\tau)} \right| + \left|\frac{\eta(\tau)}{\theta_3(\tau)} \right| +\left|\frac{\eta(\tau)}{\theta_4(\tau)} \right| \right) \ .
 \end{equation}
Here, $\tau=\tau_1 + i \tau_2$ is the complex modular parameter and $R$ is the radius of the target space circle $S^1$. The Dedekind eta function $\eta(\tau)$ and the Jacobi theta functions $\theta_i(\tau)$ are defined in Appendix~\ref{sec:ModFunc}.

\subsection{Non-Factorizable \texorpdfstring{$\boldsymbol{\mathbb{Z}_2}$}{Z2} Orbifold}

Next we turn to the $\mathbb{Z}_2$ orbifolds of non-factorizable two-dimensional tori in the absence of a background $B$-field. Recall that such tori can be described with a complex structure modulus $u$ that is a pure phase, i.e., $u=e^{i\phi}$ with $\phi\in\mathbb{R}$. We consider the $\mathbb{Z}_2$~orbifold corresponding to the involution $\iota_{\mathbb{Z}_2}$ that reflects the points on the torus along the diagonal of the primitive cell, which is the line through the origin and the lattice points $u+1$ (c.f., the solid orange diagonal in the right panel of Fig.~\ref{fig: fact and non fact}).\footnote{The $\mathbb{Z}_2$~orbifold theory corresponding to the other involution of the non-factorizable torus --- arising from the reflection along the diagonal through the points $1$ and $u$ (c.f., the dashed green line in Fig.~\ref{fig: fact and non fact}) --- is equivalent to the $\mathbb{Z}_2$~orbifold theory attributed to the involution of the first type. This can explicitly be seen by noting that the torus associated to the lattice $-\mathbb{Z} + e^{i\phi} \mathbb{Z} \simeq e^{i\phi}( \mathbb{Z} + e^{i(\pi-\phi)} \mathbb{Z})$ is equivalent to the rotated lattice $\mathbb{Z} + e^{i(\pi-\phi)} \mathbb{Z}$. As this particular conformal transformation exchanges the two described involutions, the associated toroidal $\mathbb{Z}_2$~orbifold conformal field theories are equivalent.}
This family of $\mathbb{Z}_2$~orbifold conformal field theories is parameterized by the complex structure and K\"ahler moduli of the form
\begin{equation} \label{eq:modnonfact2torus}
  u = e^{i \phi} \ , \qquad t = i \kappa \ ,\qquad \phi\in(0,\tfrac\pi2]\ ,\qquad 1\le  \kappa<\infty\ ,
\end{equation}  
which corresponds to the moduli space
\begin{equation}
  \mathcal{M}_{T^2/\mathbb{Z}_2} \simeq 
  \left((0,\tfrac\pi2]\times [1,+\infty)\right) \subset \mathcal{M}_{T^2} \ .
\end{equation}
Note that for non-factorizable tori the lattice points $1$ and $u=e^{i \phi}$ yield two circles $S^1$ of equal radii $R$ and hence equal circumferences $2\pi R$. Their radii $R$ relate to the real angular complex structure parameter $\phi$ and the real K\"ahler modulus $k$ 
as
\begin{equation}
 \kappa =\tfrac{1}{\alpha'}\lr{2\pi R}^2\sin\phi \ .
\end{equation}
Furthermore, these two circles form representatives of the homology classes generating $H_1(T^2,\mathbb{Z})$, which get exchanged by the involution~$\iota_{\mathbb{Z}_2}$. As a consequence the sum of these two homology cycles yields an invariant homology class with respect to the involution~$\iota_{\mathbb{Z}_2}$. This invariant class can be represented by the diagonal circle of $T^2$, which is the fixed-point locus of the involution.
\begin{figure}[t]
		\begin{center}
			\begin{tikzpicture}[scale=3]
			\begin{scope}[scale=1,xshift=2.5cm]
				\draw[->] (-0.2,0) -- (1.8,0) ;
				\draw[->] (0,-0.7) -- (0,0.7) ;
				\draw (0.2,0) arc (0:35:0.2);
				\node[above right] at (0.35,-0.04) {$\phi$};
				\draw[->, ultra thick] (0,0) -- (1,0) ;
				\draw[->, ultra thick] (0,0) -- ({cos(35)},{sin(35)}) ;
				\draw[very thick]  ({cos(35)},{sin(35)}) -- ({1+cos(35)},{sin(35)}) ;
				\draw[very thick]  (1,0) -- ({1+cos(35)},{sin(35)}) ;
				\draw[ultra thick, orange,->] (0,0) -- ({0.99*(1+cos(35))},{0.99*sin(35)}) ;
				\draw[ultra thick, green,->] (0,0) -- ({-cos(35)+1},{-sin(35)}) ;
				\node[below] at (1.1,0) {$1$};
				\node[above] at (0.6,0.5) {$u$};
				\node[right] at (1.2,1.2) {\framebox{$\mathbb C$}};
				\end{scope}
			\end{tikzpicture}
		\end{center}
		\caption{The lattice spanned by the orange and green vectors yields a double cover of the non-factorizable torus lattice with complex structure modulus $u$.}\label{fig: non fact double cover}
	\end{figure}
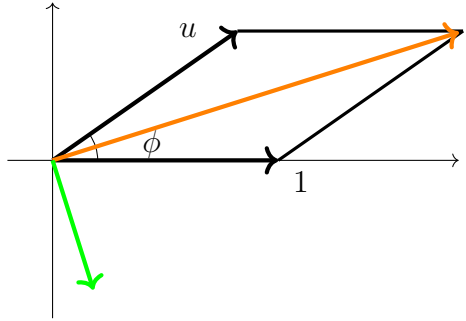
We can compute the partition function $Z_{T^2_\text{non-fac}/\mathbb{Z}_2}$ of the non-factorizable $\mathbb{Z}_2$ orbifold conformal field theory following several approaches. 

Firstly, starting from the conformal field theory of two free bosons with a two-dimensional torus as their target space characterized by the moduli~\eqref{eq:modnonfact2torus}, we orbifold this theory with the above described orbifold action to directly arrive at the partition function~$Z_{T^2_\text{non-fac}/\mathbb{Z}_2}$.

Secondly, we notice that the non-factorizable torus with moduli~\eqref{eq:modnonfact2torus} has a two-fold cover $\hat T^2$, which is a factorizable torus. The primitive cell of the non-factorizable torus lattice is spanned by generators $1$ and $e^{i\phi}$ of the lattice $\mathbb{Z} +  e^{i\phi} \mathbb{Z}$. The primitive cell of the two-fold cover is spanned by the diagonals of the original torus (see Fig. \ref{fig: non fact double cover}). One diagonal is the fixed-point circle spanned by $1+e^{i\phi}$ and the other diagonal is spanned by $1-e^{i \phi}$. Hence, we can realize the two-fold cover $\hat T^2 \simeq \mathbb{C}/(\mathbb{Z} + \hat u \mathbb{Z})$ with the complex structure modulus $\hat u$ and the complexified K\"ahler moduli $\hat t$ given by
\begin{equation}
  \hat u = \frac{1 + e^{i\phi}}{1-e^{i\phi}}= i  \cot\frac{\phi}{2} \ , \qquad
  \hat t = 2 i \kappa \ .
\end{equation}
This is done by rescaling one of the lattice generators to $1$. Thus, we can arrive at the partition function~$Z_{T^2_\text{non-fac}/\mathbb{Z}_2}$ by performing a $\mathbb{Z}_2$~shift orbifold~\cite{Wendland:2000ye} on the factorizable $\mathbb{Z}_2$~orbifold conformal field theory with the partition function~$Z_{T^2_\text{fac}/\mathbb{Z}_2}(\tau; \cot\frac\phi2,2\kappa)$ of eq.~\eqref{eq:Fact2PF}. 

Thirdly, as indicated in footnote~\ref{ft:shiftlattice} on page~\pageref{ft:shiftlattice}, the two-dimensional non-factorizable torus $T^2$ is equivalent to a two-dimensional torus $\tilde T^2\simeq \mathbb{C}/(\mathbb{Z} + \tilde u \mathbb{Z})$ with complex structure $\tilde u$ and complexified K\"ahler class $\tilde t$ given by
\begin{equation} \label{eq:modnonfact2torus_alt}
  \tilde u = \frac1{1-e^{i\phi}}=\frac12 +\frac i2 \cot\frac{\phi}{2} , \qquad
  \tilde t = i \kappa \ .
\end{equation}
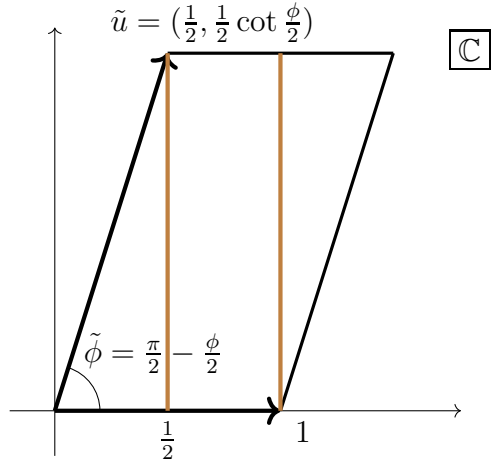
\begin{figure}
			\begin{center}
			\begin{tikzpicture}[scale=3]
				\draw[->] (-0.2,0) -- (1.8,0) ;
				\draw[->] (0,-0.2) -- (0,1.7)  ;
				\draw (0.2,0) arc (0:72.5:0.2);
                \node[above right] at (0.08,0.12) {$\tilde\phi=\frac{\pi}2-\frac{\phi}2$};
				\draw[->, ultra thick] (0,0) -- (1,0) ;
				\draw[->, ultra thick] (0,0) -- (0.5,{cot(17.5)/2}) ;
				\draw[very thick]  (1.5,{cot(17.5)/2}) -- (0.5,{cot(17.5)/2}) ;
				\draw[very thick]  (1,0) -- (1.5,{cot(17.5)/2}) ;
				\draw[ultra thick,brown] (0.5,{cot(17.5)/2}) -- (0.5,0) ;
				\draw[ultra thick,brown] (1,{cot(17.5)/2}) -- (1,0) ;
				\node[below] at (0.5,0) {$\frac{1}{2}$};
				\node[below] at (1.1,0) {$1$};
				\node[above] at (0.7,{cot(17.5)/2}) {$\tilde u=(\frac12,\frac12 \cot\frac{\phi}{2})$};
					\node[right] at (1.7,1.6) {\framebox{$\mathbb C$}};
			\end{tikzpicture}
		\end{center}
		\caption{The equivalent description of the non factorizable torus lattice $\mathbb Z+u\mathbb Z$ in the right panel of Fig. \ref{fig: fact and non fact}, via a complex structure parameter with $\text{Re}(\tilde u)=\frac12$, where $\tilde u=(1-u)^{-1}$ and $\tilde\phi=\pi/2-\phi/2$. The brown segments denote the fixed locus that corresponds to the diagonal line $u+1$ (the solid orange line in Fig. \ref{fig: fact and non fact}).}\label{fig: one half case}
	\end{figure}

In this formulation (shown in Fig. \ref{fig: one half case}), the $\mathbb{Z}_2$~orbifold acts by reflecting along the imaginary axis of the covering space $\mathbb{C}$ of $\tilde T^2$, which maps the toroidal primitive cell spanned by $1$ and $\tilde u$ to the distinct but equivalent primitive cell spanned by $1$ and $\tilde u \pm 1$.\footnote{The sign in the second generator depends on the sign of $\operatorname{Re}(\tilde u)$.} The necessary computational techniques to evaluate orbifold actions with non-invariant primitive cells are spelled out for instance in ref.~\cite{Erler:1992ki}. 

All three approaches yield the same partition function $Z_{T^2_\text{non-fac}/\mathbb{Z}_2}(\tau; \phi, \kappa)$ in terms of the moduli~\eqref{eq:modnonfact2torus}. The final result of the partition function of the $\mathbb{Z}_2$~orbifold  toroidal conformal field theory is the sum of four orbifold sectors
\begin{multline} \label{eq:sumsect}
  Z_{T^2_\text{non-fac}/\mathbb{Z}_2}(\tau; \phi, \kappa)=
  \frac12 \left( Z_{T^2_\text{non-fac}/\mathbb{Z}_2}^{(++)}(\tau; \phi, \kappa) +
  Z_{T^2_\text{non-fac}/\mathbb{Z}_2}^{(+-)}(\tau; \phi, \kappa) \right)\\[3pt]
  + \frac12 \left( Z_{T^2_\text{non-fac}/\mathbb{Z}_2}^{(-+)}(\tau; \phi, \kappa) +
  Z_{T^2_\text{non-fac}/\mathbb{Z}_2}^{(--)}(\tau; \phi, \kappa) \right) \ .
\end{multline}
These individual summands arise from traces over the Hilbert spaces  $\mathcal{H}_+$ and $\mathcal{H}_-$ of untwisted and $\mathbb{Z}_2$-twisted
states, respectively, and they are defined as 
\begin{equation}
    Z_{T^2_\text{non-fac}/\mathbb{Z}_2}^{(\pm+)} = 
    \operatorname{tr}_{\mathcal{H}_\pm} e^{2\pi\tau \hat H} \ , \quad
    Z_{T^2_\text{non-fac}/\mathbb{Z}_2}^{(\pm -)} = 
    \operatorname{tr}_{\mathcal{H}_\pm} g \, e^{2\pi\tau \hat H} \ ,
\end{equation}
where $\hat H$ is the Hamiltonian of the conformal field theory, $g$ is the generator of the $\mathbb{Z}_2$~orbifold group, which acts on the states in the Hilbert spaces $\mathcal{H}_\pm$. Thus, the first line in the expansion~\eqref{eq:sumsect} projects onto untwisted $\mathbb{Z}_2$~invariant states and the second line corresponds to a sum of twisted $\mathbb{Z}_2$~invariant states. For more details on such orbifold constructions see, e.g., ref.~\cite{Blumenhagen:2013fgp}. Summing up all these contributions is conveniently expressed in terms of the circle partition function $Z_{S^1}$ \eqref{eq:PFS1}, which takes the form
\begin{align}\label{eq: non fact Z2 partition function}
&Z_{T^2_\text{non-fac}/\mathbb{Z}_2}(\tau; \phi, \kappa)=
\frac14 \left|\frac{\theta_2(\tau)}{\eta(\tau)} \right|^2 
Z_{S^1}(2\tau;\tfrac{R_1}{\sqrt{2}})\,Z_{S^1}(2\tau;\tfrac{R_2}{\sqrt{2}})\\
&\qquad+\frac14  \left|\frac{\theta_4(\tau)}{\eta(\tau)} \right|^2
Z_{S^1}(\tfrac\tau2;\tfrac{R_1}{\sqrt{2}})\, Z_{S^1}(\tfrac\tau2;\tfrac{R_2}{\sqrt{2}})
+\frac14  \left|\frac{\theta_3(\tau)}{\eta(\tau)} \right|^2
Z_{S^1}(\tfrac{\tau+1}2;\tfrac{R_1}{\sqrt{2}})\,
Z_{S^1}(\tfrac{\tau+1}2;\tfrac{R_2}{\sqrt{2}})\nn\\
&\qquad-\frac12 Z_{S^1}(\tau; R_1)\, Z_{S^1}(\tau; \tfrac{R_2}2) 
- \frac12Z_{S^1}(\tau; \tfrac{R_1}2)\, Z_{S^1}(\tau; R_2)\nn \\
&\qquad+\frac12 Z_{S^1}(2\tau; \tfrac{R_1}{\sqrt{2}}) +
\frac12 Z_{S^1}(\tfrac\tau2; \tfrac{R_1}{\sqrt{2}}) +
\frac12 Z_{S^1}(\tfrac{\tau+1}2; \tfrac{R_1}{\sqrt{2}}) \ ,\nn
\end{align}
where
\begin{equation} \label{eq:R1R2rel}
 2\pi  R_1 = \sqrt{2\alpha'\kappa \cot\frac\phi2} \ , \qquad 2\pi  R_2 =\sqrt{ 2\alpha' \kappa \tan\frac\phi2} \ .
\end{equation}
$R_1$ corresponds to the orange segment of Fig. \ref{fig: non fact double cover} and $R_2$ to the green segment of the same figure.\footnote{$R_1$ and $R_2$ can be also written as $\frac{R_1}{\sqrt{2}} = 2\pi R\sqrt{\lr{1+\cos\phi}}$ and $\frac{R_2}{\sqrt{2}} = 2\pi R\sqrt{\lr{1-\cos\phi}}\ .$} The first three lines of eq. \eqref{eq: non fact Z2 partition function} correspond to the summand~$Z_{T^2_\text{non-fac}/\mathbb{Z}_2}^{(++)}$ of the untwisted sector with no insertion of the orbifold generator, and the last line contains contributions from the untwisted sector with insertion $Z_{T^2_\text{non-fac}/\mathbb{Z}_2}^{(+-)}$, and the twisted sector without insertion $Z_{T^2_\text{non-fac}/\mathbb{Z}_2}^{(-+)}$ and the twisted sector with insertion $Z_{T^2_\text{non-fac}/\mathbb{Z}_2}^{(--)}$. Note that in the last line the moduli dependence is only on $R_1$, i.e. the length of the orbifold fixed point locus.

If for $\phi=0$ the two diagonal radii $R_1$ and $R_2$ become equal (c.f., eq.~\eqref{eq:R1R2rel}), the two-dimensional target space torus factorizes into two circles both of radius~$\frac1{\sqrt{2}}R_1 =\frac1{\sqrt{2}}R_2$. However, the $\mathbb{Z}_2$~orbifold acts differently than for the factorizable tori studied in subsection~\ref{sec:factorb}, as it exchanges the two equally-sized circles. In the terminology of ref.~\cite{Kames-King:2023fpa} the conformal field theory simplifies to the $S_2\simeq\mathbb{Z}_2$~orbifold conformal field theory of the product of two circles. Namely, for $\frac1{\sqrt{2}}R_1 =\frac1{\sqrt{2}}R_2$ the partition function~\eqref{eq: non fact Z2 partition function} becomes 
\begin{multline}
  Z_{T^2_\text{non-fac}/\mathbb{Z}_2}(\tau; 0, \tfrac{2\pi^2 R_1^2}{\alpha'})= \\\frac12
  Z_{S^1}(\tau;\tfrac{R_1}{\sqrt{2}})^2 
  + \frac12 \left( Z_{S^1}(2\tau; \tfrac{R_1}{\sqrt{2}}) +
 Z_{S^1}(\tfrac\tau2; \tfrac{R_1}{\sqrt{2}}) +
 Z_{S^1}(\tfrac{\tau+1}2; \tfrac{R_1}{\sqrt{2}})\right) \ ,
\end{multline}
which is indeed in agreement with the partition function of the $S_2$ symmetric orbifold conformal field theories studied ref.~\cite{Kames-King:2023fpa}, where the $S_2$ permutes the two equally sized target space circles $S^1$.

\subsection{Two-Dimensional Toroidal \texorpdfstring{$\boldsymbol{\mathbb{Z}_2}$}{Z2} Orbifolds with \texorpdfstring{$\boldsymbol B$}{B}-Field}
For $\mathbb{Z}_2$ orbifold theories with a non-vanishing $B$-field, we distinguish again between factorizable and non-factorizable two-dimensional tori.

\paragraph{Factorizable orbifold:}Factorizable two-dimensional tori with a non-vanishing $B$-field admit the discussed $\mathbb{Z}_2$~orbifold action if their moduli are constrained to
\begin{equation} \label{eq:modnonfac_B}
  u = i c \ , \qquad t = \frac12 + i \kappa\ ,\qquad1\leq c<\infty\ ,\qquad\frac12\leq\kappa<\infty\ ,
\end{equation}
in terms of the real parameter $c$ and the real K\"ahler modulus $\kappa$. This complexified K\"ahler modulus~$t$ is invariant with respect to the $\mathbb{Z}_2$~orbifold action because $\iota_{\mathbb{Z}_2}^*$ flips the sign of the $B$-field $B$, which for the non-vanishing value $b=\frac{1}{2}$ remains invariant due to the periodicity $b\sim b+1$. Note that mirror symmetry --- which exchanges the complex structure modulus $u$ and the K\"ahler modulus $t$ --- maps the moduli~\eqref{eq:modnonfac_B} to the mirror dual moduli
\begin{equation} \label{eq:modmirrordual}
  \tilde u = \frac12 + i \kappa \ , \qquad
\tilde t = i c \ ,
\end{equation}
which are the moduli of the description in terms of a non-factorizable torus with vanishing $B$-field as given in eq.~\eqref{eq:modnonfact2torus_alt}. As a result, we readily obtain the partition function $Z_{T^2_{\text{fac}/\mathbb{Z}_2}}^{b=1/2}$ from the partition function $Z_{T^2_\text{non-fac}/\mathbb{Z}_2}$ via mirror symmetry. To do so, we map the complex structure modulus of eq.~\eqref{eq:modmirrordual} to the unit circle -- as in footnote~\ref{ft:shiftlattice}. This gives
\begin{equation}
    \tilde u \mapsto \tilde u=\frac{4\kappa^2-1}{4\kappa^2+1}+i\frac{4\kappa}{4\kappa^2+1}\ . 
\end{equation}
Taking this into account, we get
\begin{align} \label{eq:MirrorCorr}
    Z_{T^2_\text{fac}/\mathbb{Z}_2}^{b=1/2}(\tau; c, \kappa) =Z_{T^2_\text{non-fac}/\mathbb{Z}_2}(\tau;  \arctan\lr{\tfrac{4\kappa}{4\kappa^2-1}},c) \ .
\end{align}  

\paragraph{Non-Factorizable orbifold:} 
Finally we consider non-factorizable two-dimensional tori with non-vanishing $B$-field. In this case, the complex structure modulus $u$ and the complexified K\"ahler modulus $t$ are both taken to be phases. However, we have not been able to express the partition functions in a convenient form as in eq.~\eqref{eq: non fact Z2 partition function}, which, as we will see soon in subsection \ref{subsec_avg}, makes the averaging rather straightforward.

\subsection{Partition Functions and Siegel--Narain Theta Functions}\label{sec: section 2 thetas}
In order to calculate efficiently the ensemble average of conformal field theories arising from toroidal target spaces as developed in ref.~\cite{Dong:2021}, it is convenient to express the partition functions in terms of the Siegel--Narain theta functions \cite{Siegel44,siegel1955lectures,Dong:2021}. Let $\Omega$ be a symmetric $2N\times 2N$ matrix of signature $(N,N)$, such that $2\Omega$ has integral entries and even entries on the diagonal.\footnote{In refs.~\cite{Siegel44,siegel1955lectures}, Siegel constructs theta functions for symmetric non-degenerate pairings $\Omega$ with arbitary signature $(r,s)$.} Moreover, let $H$ be a symmetric positive definite real $2N\times 2N$ matrix obeying
\begin{equation} \label{eq:Hrelation}
    H \Omega^{-1} H = \Omega \ .
\end{equation}
Then the Siegel--Narain theta functions in terms of $\Omega$ and $H$ are defined as \cite{Siegel44,siegel1955lectures,Dong:2021}
\begin{equation}
 \Theta_{H,\Omega}(a,b,\tau)
  =   \sum_{m\in \mathbb{Z}^{2N}} 
  e^{-2 \pi \operatorname{Im}(\tau)\, (m+b)^T H (m+b) + 2 \pi i \operatorname{Re}(\tau)\, (m+b)^T \Omega (m+b) 
  - 4 \pi i \operatorname{Re}(\tau)\, a^T \Omega (m+\frac12 b)} \ .
\end{equation}
with the twist vectors $a, b \in \mathbb{R}^{2N}$ and the modular parameter $\tau$ in the upper half-plane~$\mathcal{H}$. Note that the positive definiteness of the matrix $H$ ensures that the summation over $m\in\mathbb{Z}^{2N}$ converges in the definition of the theta function $\Theta_{H,\Omega}$. The Siegel--Narain theta functions of this work are all defined with respect to the pairing
\begin{equation} \label{eq:defOmega}
   \Omega = \frac12 \begin{pmatrix} 0 & \mathbf{1}_{N\times N} \\ \mathbf{1}_{N\times N} & 0 \end{pmatrix} \ .
\end{equation}   
Hence, for ease of notation we only refer to the positive definite $2N\times 2N$~matrix~$H$ in the expression for the Siegel--Narain theta functions, i.e.,
\begin{equation} \label{eq:SNtheta}
    \Theta_{H}(a,b,\tau) \equiv  \Theta_{H,\Omega}(a,b,\tau) \ .
\end{equation}
In terms of the Siegel--Narain theta functions defined in eq.~\eqref{eq:SNtheta} the partition function $Z_{T^{2\ell}}(\tau;G,B)$ of the toroidal conformal field theory with the target space torus $T^{2\ell}$ becomes (see, e.g., ref.~\cite{Blumenhagen:2013fgp})
\begin{equation} \label{eq:Ztor2l}
   Z_{T^{2\ell}}(\tau; G, B) = 
   \frac{1}{|\eta(\tau)|^{4\ell}} \Theta_{H(G,B)}(0,0,\tau) \ , 
\end{equation}   
with the real positive definite $2\ell\times2\ell$-matrix $H(G,B)$ (which obeys the relation~\eqref{eq:Hrelation})
\begin{equation} \label{eq:Hrel}
    H(G,B) = {\scriptstyle\begin{pmatrix}
        \frac{\alpha'}{2}G^{-1} & \frac12 G^{-1}B\\[3pt]
        -\frac12 BG^{-1} & \frac{1}{2\alpha'}\lr{G-BG^{-1}B}
    \end{pmatrix}}\ ,
\end{equation}
with inverse
\begin{equation}
     H^{-1}(G,B) = {\scriptstyle\begin{pmatrix}
        \frac{2}{\alpha'}\lr{G-BG^{-1}B} & -2 B G^{-1}\\[3pt]
        2G^{-1}B & 2\alpha' G^{-1}
    \end{pmatrix}} \ .
\end{equation}

In this subsection we consider the contribution to the partition functions of the toroidal $\mathbb Z_2$ orbifolds coming from the untwisted sector, and with no insertion of the involution element. We refer to this contribution as $Z_{T^2_\text{non-fac}/\mathbb{Z}_2}^{(+,+)}(\tau)$. We express $Z_{T^2_\text{non-fac}/\mathbb{Z}_2}^{(+,+)}(\tau)$ in terms of the Siegel-Narain theta functions defined in eq.~\eqref{eq:SNtheta}. The resulting expressions will in particular be useful when we compute the average partition function over the moduli space for higher dimensional target spaces in section~\ref{sec:ObCFTTD}, following the methods of ref.~\cite{Dong:2021}.

For the factorizable $\iota_{\mathbb Z_2}$ involution, the contribution from the untwisted sector with no insertions to eq. \eqref{eq:Fact2PF} is of the form:
\begin{equation}\label{eq: 1d theta fact}
Z_{T^2_\text{fac}/\mathbb{Z}_2}^{(+,+)}(\tau;c,\kappa)
=  \frac{1}{\ \abs{\eta(\tau)}^{4}}\;
\Theta_{H(g)}\lr{0,0,\tau}\Theta_{H(\wt g)}\lr{0,0,\tau }
\end{equation}
where $g,\wt  g$ are the metrics of the two circles.

For the non-factorizable involution \eqref{eq: non fact Z2 partition function}, we find
\begin{equation}\label{eq: circle circle non fact theta pf}
 Z^{(+,+)}_{T^{2}_\text{non-fac}}(\tau;\phi,\kappa) = 
   \frac{1}{\abs{\eta(\tau)}^{4}}\sum_{\Delta \in \{0,1\}^{2}} 
   \Theta_{h}(0,\tfrac12\Delta,2\tau)\ \Theta_{{\wt h}}(0,\tfrac12\Delta,2\tau) \ .
 	\end{equation}
 In fact, one can show that the above expressions are modular invariant using the properties of the twisted theta functions under modular transformations. We shall generalize these formulas in section \ref{sec:ObCFTTD}, where we study higher dimensional toroidal target spaces.

\subsection{Ensembles of Two-Dimensional Toroidal \texorpdfstring{$\boldsymbol{\mathbb{Z}_2}$}{Z2} Orbifolds}\label{subsec_avg}
The moduli space $\mathcal{M}_{T^2}$ of the conformal field theory of two free bosons with a two-dimensional torus $T^2$ as its target space may be locally parameterized in terms of the complex structure modulus~$u$ and the complexified K\"ahler modulus $t$. The two-point correlators of the marginal operators define the Weil--Petersson metric on the moduli space $\mathcal{M}_{T^2}$ \cite{Zamolodchikov:1986gt}, which for the toroidal conformal field theory is locally a product of two two-dimensional hyperbolic spaces with the metric (up to a constant pre-factor)
\begin{equation} \label{eq:MetModT2}
 \dd s^2 = \frac{\dd u\,\dd\bar u}{(\operatorname{Im}{u})^2} + \frac{\dd t \, \dd\bar t}{(\operatorname{Im} t)^2} \ .
\end{equation}
This comes from the fact that the moduli space of $T^2$ compactifications is locally a product of two copies of the fundamental domain (subset of the upper half plane). One may explicitly arrive at the above expression (up to normalization) by calculating the Zamolodchikov metric
    \begin{align}\label{eq: Zam metric def}
    \dd s^2=G^{mp}G^{nq}\lr{\mathrm{d}G_{mn}\mathrm{d}G_{pq}+\mathrm{d}B_{mn}\mathrm{d}B_{pq}}
\end{align} and plugging in the metric and the $B$-field in terms of the complex structure and complexified K\"ahler moduli $u,t$. 
The moduli spaces $\mathcal{M}_{T^2/\mathbb{Z}_2}$ of the analyzed $\mathbb{Z}_2$~orbifold conformal field theories are subspaces of $\mathcal{M}_{T^2}$. The Weil--Petersson metric of the moduli space $\mathcal{M}_{T^2/\mathbb{Z}_2}$ is the induced metric from the metric~\eqref{eq:MetModT2}, because the exactly marginal operators in the untwisted sector of the orbifold theories have the same two-point correlation functions as in the unorbifolded theory.

\subsubsection{Factorizable $\boldsymbol{\mathbb{Z}_2}$ orbifold with vanishing $\boldsymbol B$-field}
The parameters of this $\mathbb{Z}_2$~orbifold are the two real moduli $(c, \kappa)$ of eq.~\eqref{eq:modfact2torus}. A fundamental domain of the moduli $(c,\kappa)$ reads
\begin{equation}
   (c, \kappa) \in [1, \infty) \times [1,\infty) \ .
\end{equation}  
The measure of the moduli induced from eq.~\eqref{eq:MetModT2} becomes
\begin{equation}
  \dd s^2 = \left(\frac{\dd c}{c}\right)^2 + \left(\frac{\dd\kappa}{\kappa} \right)^2 \ .
\end{equation}  
The volume of the moduli space of this $\mathbb{Z}_2$~orbifold theory is logarithmically divergent. Note that for this class of conformal field theories, the ensemble average over the entire moduli space is divergent as well \cite{Maloney:2020nni}. Nevertheless, we can still study ensemble averages over measurable subsets of the moduli space by, for instance, regularizing the integral with cut off $\Lambda \gg 1$ for large (and small) values of the moduli  $c$ and $k$. Using the partition function \eqref{eq:Fact2PF}, we arrive at
\begin{align}
     \left\langle Z_{T^2_\text{fac}/\mathbb{Z}_2}(\tau) \right\rangle_\text{reg} 
     &=\frac{1}{\lr{\text{Vol}_{S^1,\text{reg}}}^2} \int_{1}^\Lambda \frac{\dd c}c \int_{1}^\Lambda \frac{\dd\kappa}\kappa \, \bigg\{ \frac12 Z_{S^1}(\tau; R_1(c,\kappa)) \, Z_{S^1}(\tau; R_2(c,\kappa))+\\ &\qquad\;\;\left( \left|\frac{\eta(\tau)}{\theta_2(\tau)} \right| + \left|\frac{\eta(\tau)}{\theta_3(\tau)} \right| +\left|\frac{\eta(\tau)}{\theta_4(\tau)} \right| \right)
     Z_{S^1}(\tau; R_2(c,\kappa))\bigg\}\ ,\nn 
\end{align}
in terms of the regularized moduli space volume
\begin{equation}
    \text{Vol}_{S^1,\text{reg}} :=\int_1^\Lambda\frac{\dd x}{x} = \log \Lambda \ .
\end{equation}
Defining for the partition function $Z_{S^1}$ in eq.~\eqref{eq:PFS1} the regularized ensemble average by
\begin{equation} \label{eq:CircAverage}
    \left\langle Z_{S^1}(\tau) \right\rangle_\text{reg} 
    = \frac{1}{\text{Vol}_{S^1,\text{reg}}}\int_{1}^\Lambda \frac{\dd R}{R} \, Z_{S^1}(\tau; R) \ ,
\end{equation}
we express the regularized ensemble average $\left\langle Z_{T^2_\text{fac}/\mathbb{Z}_2}(\tau) \right\rangle_\text{reg}$ as: 
\begin{align}\label{eq average of fact}
    \left\langle Z_{T^2_\text{fac}/\mathbb{Z}_2}(\tau) \right\rangle_\text{reg}
    &= \frac12\left\langle Z_{S^1}(\tau) \right\rangle_\text{reg}^2 +  \left( \left|\frac{\eta(\tau)}{\theta_2(\tau)} \right| + \left|\frac{\eta(\tau)}{\theta_3(\tau)} \right| +\left|\frac{\eta(\tau)}{\theta_4(\tau)} \right| \right) 
    \left\langle Z_{S^1}(\tau) \right\rangle_\text{reg} \ .\nn
\end{align}
The first summand is simply the (regularized) ensemble average of the tensor product of two circular conformal field theories, whereas the second contribution comes from the circular fixed-point loci of the $\mathbb{Z}_2$ orbifold. 

\subsubsection{Non-factorizable $\boldsymbol{\mathbb{Z}_2}$ orbifold with vanishing $\boldsymbol B$-field}
The moduli space of this class of toroidal $\mathbb{Z}_2$ orbifold is parametrized by the angular complex structure modulus~$\phi$ and the real K\"ahler modulus $\kappa$ in the range \eqref{eq:modnonfact2torus}.
From eq.~\eqref{eq:MetModT2} we arrive at the induced moduli space metric
\begin{equation}
  \dd s^2 = \left( \frac{\dd\phi}{\sin \phi}\right)^2 + \left( \frac{\dd\kappa}{\kappa} \right)^2 \ .
\end{equation}
The volume of the moduli space exhibits logarithmic divergences as $\phi$ and $\kappa$ approach zero and $+\infty$, respectively. Thus, we define the regularized ensemble average
\begin{align} \label{eq:T2nonfac_reg1}
   \left\langle Z_{T^2_\text{non-fac}/\mathbb{Z}_2}(\tau) \right\rangle_\text{reg}
   = \frac{1}{V_{\text{reg}}}\int_\delta^{\pi-\delta} \frac{\dd\phi}{\sin\phi} \int_{\frac1\Lambda}^\Lambda \frac{\dd\kappa}\kappa \; Z_{T^2_\text{non-fac}/\mathbb{Z}_2}(\tau; \phi, \kappa) \ .
\end{align}
Here the ensemble average is regularized by introducing a small positive angle $\delta$ for the angular variable $\phi$ and a large value $\Lambda$ for the K\"ahler modulus $\kappa$.
We integrate over a 4-fold cover of the moduli space \eqref{eq:modnonfact2torus}, which is normalized by
\begin{equation}
    V_{\text{reg}} = \int_\delta^{\pi-\delta} \frac{\dd\phi}{\sin\phi} \int_{\frac1\Lambda}^\Lambda \frac{\dd\kappa}\kappa \ .
\end{equation}
Upon expressing the moduli in terms of $R_1$ and $R_2$ defined in eq.~\eqref{eq:R1R2rel}, we obtain for the regularized ensemble average
\begin{multline}
   \left\langle Z_{T^2_\text{non-fac}/\mathbb{Z}_2}(\tau) \right\rangle_\text{reg} \\
   =\frac{1}{4\lr{\text{Vol}_{S^1,\text{reg}}}^2} \int_{\frac1{\Lambda_1}}^{\Lambda_1} \frac{\dd R_1}{R_1} \int_{\frac1{\Lambda_2}}^{\Lambda_2} \frac{\dd R_2}{R_2} \,
   Z_{T^2_\text{non-fac}/\mathbb{Z}_2}\left(\tau; 2 \operatorname{arctan}\left(\tfrac{R_2}{R_1}\right), \tfrac{2\pi^2}{\alpha'} R_1 R_2\right) \ ,
\end{multline}
where the large cut-offs $\Lambda_1$ and $\Lambda_2$ arise from the transformation the regulators $\delta$ and $\Lambda$ in eq.~\eqref{eq:T2nonfac_reg1} and the normalization factor is given by $V_{\text{reg}}=8\text{Vol}_{S^1,\text{reg}}$. Expressed in terms of the circular ensemble average~\eqref{eq:CircAverage}, this (regularized) ensemble average becomes 
\begin{multline} \label{eq:nonfact_av}
   \left\langle Z_{T^2_\text{non-fac}/\mathbb{Z}_2}(\tau) \right\rangle_\text{reg}
   =\\= \frac14 \left( \left|\frac{\theta_2(\tau)}{\eta(\tau)} \right|^2 \left\langle Z_{S^1}(2\tau) \right\rangle_\text{reg}^2  
   +  \left|\frac{\theta_4(\tau)}{\eta(\tau)} \right|^2  \left\langle Z_{S^1}(\tfrac\tau2) \right\rangle_\text{reg}^2 
   +  \left|\frac{\theta_3(\tau)}{\eta(\tau)} \right|^2  \left\langle Z_{S^1}(\tfrac{\tau+1}2) \right\rangle_\text{reg}^2  \right) \\
   - \left\langle Z_{S^1}(\tau) \right\rangle_\text{reg}^2
    +\frac12 \left(  \left\langle Z_{S^1}(2\tau) \right\rangle_\text{reg} + \left\langle Z_{S^1}(\tfrac\tau2) \right\rangle_\text{reg} 
   + \left\langle  Z_{S^1}(\tfrac{\tau+1}2) \right\rangle_\text{reg} \right) 
   \ .
\end{multline} 
Note that the individual terms assembled in brackets in eq.~\eqref{eq:nonfact_av} form modular invariant combinations with respect to the modular group acting on the worldsheet modular parameter $\tau$. 

\subsubsection{$\boldsymbol{\mathbb{Z}_2}$ orbifold with non-vanishing $\boldsymbol B$-field}
\paragraph{Factorizable orbifold:} Due to the relation~\eqref{eq:MirrorCorr} the (regularized) ensemble average of the factorizable $\mathbb{Z}_2$ orbifold theories with the $B$-field background $\text{Re}(t)=\frac12$ are identical to the regularized ensemble averages of the non-factorizable $\mathbb{Z}_2$~orbifold theories with vanishing background $B$-field. That is to say, upon averaging over the moduli space given in eq.~\eqref{eq:modnonfac_B},
we arrive for the ensemble averages at the mirror correspondence
\begin{equation}
  \left\langle Z_{T^2_\text{fac}/\mathbb{Z}_2}^{b=1/2}(\tau) \right\rangle_\text{reg} = \left\langle Z_{T^2_\text{non-fac}/\mathbb{Z}_2}(\tau) \right\rangle_\text{reg} \ ,
\end{equation}  
where the right-hand side is explicitly given in eq.~\eqref{eq:nonfact_av}.

\paragraph{Non-factorizable orbifold:}The moduli space is parametrized in terms of the angular complex structure variable $\phi$ and the real K\"ahler modulus $k$ in the range
\begin{equation}
  \phi \in (0,\tfrac\pi2] \ , \qquad k \in [\tfrac12,+\infty) \ .
\end{equation}  
The partition function for this orbifold cannot be simply obtained by a duality argument. While this is an interesting case to consider, we do not pursue this further in this work.

\section{Orbifold CFTs from \texorpdfstring{$\boldsymbol D$}{D}-Dimensional Tori \texorpdfstring{$\boldsymbol{T^D}$}{TD}} \label{sec:ObCFTTD} 
In order to obtain finite ensemble averages of moduli spaces of finite volume from toroidal orbifold conformal field theories, it is necessary to consider toroidal target spaces of higher dimensions as in refs.~\cite{Maloney:2020nni,Afkhami-Jeddi:2020ezh}. Therefore, we now generalize the conformal field theories analyzed in Section~\ref{sec:ObCFTT2} to higher dimensional toroidal conformal field theories and orbifolds thereof. Namely, we now consider $D$ free bosons for $D\ge 6$ with periodic boundary conditions parametrizing the target space torus $T^D$. The $\mathbb{Z}_2$~orbifold action is again characterized by an involution $\iota_{\mathbb{Z}_2}: T^D \to T^D$ together with the induced action via the pull-back $\iota^*_{\mathbb{Z}_2}$ acting on the toroidal metric and the two-form $B$-field.

In ref.~\cite{Benjamin:2021wzr} $\mathbb{Z}_2$~orbifolds of toroidal conformal field theories and their ensemble averages are studied for involutions $\iota_{\mathbb{Z}_2}$ that invert all directions parametrized by the bosons. In this section we extend this class of $\mathbb{Z}_2$~orbifold toroidal conformal field theories and calculate their ensemble averages. 

\subsection{Factorizable Toroidal \texorpdfstring{$\boldsymbol{\mathbb{Z}_2}$}{Z2}~Orbifold CFTs} \label{sec:FacTorOrbTD}
Our first class of theories arises from factorizable tori $T^D\simeq T^{\ell} \times T^{m}$ equipped with a product metric and a block diagonal $B$-field. Such tori admit a $\mathbb{Z}_2$~orbifold action resulting from the involution~$\iota_{\mathbb{Z}_2}$, which is defined via the action 
\begin{equation} \label{eq:orbactTDfac}
  \iota_{\mathbb{Z}_2}: \ (x,y) \mapsto (-x, y) \ ,
\end{equation}
where $(x,y)$ are the coordinates on the universal covering space $\mathbb{R}^{\ell} \times \mathbb{R}^{m}$ of the torus $T^\ell \times T^m$. Moreover, the block-diagonal metric and the block-diagonal $B$-field are invariant with respect to the pull-back~$\iota_{\mathbb{Z}_2}^*$. We denote the toroidal $\mathbb{Z}_2$~orbifold resulting from the involution~$\iota_{\mathbb{Z}_2}$ by
\begin{equation} \label{eq:TDfac}
  T^D_\text{fac}/{\mathbb{Z}_2} \simeq T^{\ell}/{\mathbb{Z}_2} \times T^{m} \ , 
  \quad D = \ell +  m \ , \quad \ell,m\ge 3 \ .
\end{equation}

As this class of conformal field theories simply arises from a product of the type of toroidal $\mathbb{Z}_2$~orbifold conformal field theories studied in ref.~\cite{Benjamin:2021wzr} and a toroidal conformal field theory with target~$T^{2\ell}$ considered in ref.~\cite{Maloney:2020nni}, the partition function is a product, which readily generalizes expression~\eqref{eq:Fact2PF} as
\begin{equation} \label{eq:TDfacOrbPar}
  Z_{T^D_{\text{fac}}/{\mathbb{Z}_2}}(\tau; \mathfrak{l}, \mathfrak{m}) 
  = Z_{T^{\ell}/{\mathbb{Z}_2}}(\tau; \mathfrak{l})  Z_{T^{m}}(\tau; \mathfrak{m}))\ .
\end{equation}
The second factor is the partition function of an $m$-dimensional toroidal conformal field theory with the moduli $\mathfrak{m}$ \cite{Blumenhagen:2013fgp,Maloney:2020nni} (see Appendix~\ref{sec:ModFunc}). The first factor reads \cite{Benjamin:2021wzr} 
\begin{equation}\label{eq fact pf higher d}
   Z_{T^{\ell}/{\mathbb{Z}_2}}(\tau; \mathfrak{l}) = \frac12 
   \left( Z_{T^{\ell}}(\tau; \mathfrak{l}) + 2^{\ell} \left[ 
   \left| \frac{\eta(\tau)}{\theta_2(\tau)}\right|^{\ell} 
   + \left| \frac{\eta(\tau)}{\theta_3(\tau)}\right|^{\ell} 
   + \left| \frac{\eta(\tau)}{\theta_4(\tau)}\right|^{\ell} \right]\right) \ .
\end{equation}
Here $Z_{T^{\ell}}(\tau; \mathfrak{l})$ is the partition function of the unorbifolded $\ell$-dimensional toroidal conformal field theory with moduli $\mathfrak{l}$, whereas the remaining terms relate to $\mathbb{Z}_2$~orbifold contributions \cite{Benjamin:2021wzr}.

For the considered toroidal orbifolds~\eqref{eq:TDfac} with their block-diagonal metric and with their block-diagonal $B$-field the product structure of the partition function~\eqref{eq:TDfacOrbPar} not only prevails for a particular choice of the moduli $\mathfrak{m}$ and $\mathfrak{l}$, but also holds globally over the entire moduli space $\mathcal{M}_{T^D_\text{fac}/\mathbb{Z}_2}$ of the underlying family of conformal field theories, i.e., 
\begin{equation} \label{eq:ProdModfac}
  \mathcal{M}_{T^D_\text{fac}/\mathbb{Z}_2} 
  = \mathcal{M}_{T^\ell}  \times \mathcal{M}_{T^m\vphantom{\ell}} \ .
\end{equation}
Here $\mathcal{M}_{T^d}$ denotes the moduli space of conformal field theories of a $d$-dimensional torus $T^d$.\footnote{As the orbifold action~$\mathbb{Z}_2$ reflects all directions of the first factor $T^\ell$, the orbifold action is well-defined for any point in the moduli space $\mathcal{M}_{T^\ell}$. Therefore, the moduli spaces $\mathcal{M}_{T^\ell}$ and $\mathcal{M}_{T^\ell/\mathbb{Z}_2}$ are identical.} Therefore, the resulting ensemble average of this class of $\mathbb{Z}_2$~orbifold conformal field theories readily becomes
\begin{equation} \label{eq:EAfacTD}
\begin{aligned}
  \left\langle Z_{T^D_{\text{fac}}/{\mathbb{Z}_2}}(\tau) \right\rangle &=
  \int_{{\mathcal{M}_{T^\ell}}{\times}{\mathcal{M}_{T^{m\phantom{\ell}}}}} \dd\mu(\mathfrak{l},\mathfrak{m}) \, Z_{T^D_{\text{fac}}/{\mathbb{Z}_2}}(\tau; \mathfrak{l}, \mathfrak{m}) \\
  &= \left( \int_{\mathcal{M}_{T^\ell}} \dd\mu(\mathfrak{l})\,Z_{T^{\ell}/{\mathbb{Z}_2}}(\tau; \mathfrak{l})  \right) 
  \left(\int_{\mathcal{M}_{T^m}} \dd\mu(\mathfrak{m})  \,Z_{T^{m}}(\tau; \mathfrak{m})) \right) \\[1ex]
  &= \left\langle \strut Z_{T^{\ell}/\mathbb{Z}_2}(\tau) \right\rangle
      \left\langle \strut Z_{T^{m}}(\tau) \right\rangle \ ,
\end{aligned}  
\end{equation}  
and factorizes into a product of ensemble averages. Here  $d\mu(\mathfrak{l},\mathfrak{m})$ is the measure of the moduli space $\mathcal{M}_{T^D_\text{fac}/\mathbb{Z}_2}$.  It factors into the measures $d\mu(\mathfrak{l})$ and $d\mu(\mathfrak{m})$ for the moduli spaces of toroidal conformal field theories, which (for arbitrary $d$-dimensional tori) is normalized to 
\begin{equation}
    \operatorname{Vol}_{T^d} = \int_{\mathcal{M}_{T^d}}  \dd \mu(\mathfrak{d}) = 1 \ .
 \end{equation}   
The ensemble average of a $d$-dimensional toroidal conformal field theory is calculated as in ref.~\cite{Maloney:2020nni} and reads 
\begin{equation} \label{eq:AvZTd}
  \left\langle \strut Z_{T^{d}}(\tau) \right\rangle 
  =  \int_{\mathcal{M}_{T^d}}  \dd \mu(\mathfrak{d}) Z_{T^{d}}(\tau;\mathfrak{d})
  = \frac{E_{\nicefrac d2}(\tau)}{\text{Im}(\tau)^{\frac{d}2}\abs{\eta(\tau)}^{2d}}    \ ,  \quad d\ge 3 \ .
\end{equation}
As a result and together with eq.~\eqref{eq:TDfacOrbPar} the ensemble average of the partition function~\eqref{eq:EAfacTD} becomes
\begin{multline} \label{eq:AvOrb}
  \left\langle Z_{T^D_{\text{fac}}/{\mathbb{Z}_2}}(\tau) \right\rangle 
  = \frac12 
 \frac{ E_{\nicefrac \ell2}(\tau) E_{\nicefrac m2}(\tau)}{\text{Im}(\tau)^{\frac{\ell+m}2}\abs{\eta(\tau)}^{2(\ell+m)}}  \\
  + 2^{\ell-1} \left[ 
   \left| \frac{\eta(\tau)}{\theta_2(\tau)}\right|^{\ell} 
   + \left| \frac{\eta(\tau)}{\theta_3(\tau)}\right|^{\ell} 
   + \left| \frac{\eta(\tau)}{\theta_4(\tau)}\right|^{\ell} \right] 
   \frac{E_{\nicefrac m2}(\tau)}{\text{Im}(\tau)^{\frac{m}2}\abs{\eta(\tau)}^{2 m}} \ .
\end{multline}
Here and in eq.~\eqref{eq:AvZTd}, $E_s(\tau)$ denotes the real analytic Eisenstein series
\begin{equation} \label{eq:defEisenstein}
  E_s(\tau) = \frac12 \sum_{\genfrac{}{}{0pt}{2}{c,d\in\mathbb{Z}}{  (c,d)=1}} \frac{\operatorname{Im}(\tau)^s}{| c \tau + d|^{2s}} \ ,
\end{equation} 
where in the summation $(c,d)$ denotes the greatest common divisor of the integers $c$ and $d$. The real analytic Eisenstein series $E_s(\tau)$ is a modular function in $\tau$ that is defined for $s\in \mathbb{C}$ with $\operatorname{Re}(s) > \frac12$. For more details on the real Eisenstein series see for instance refs.~\cite{10.1007/978-3-662-00734-1_10, terras2013harmonic} and Appendix~\ref{sec:ModFunc}.

\subsection{Non-Factorizable Toroidal \texorpdfstring{$\boldsymbol{\mathbb{Z}_2}$}{Z2}~Orbifold CFTs}
A systematic classification of all $\mathbb{Z}_2$~orbifolds of toroidal conformal field theories is beyond the scope of this work. Instead we focus on an interesting class of toroidal $\mathbb{Z}_2$~orbifolds that generalizes the factorizable $\mathbb{Z}_2$~orbifolds of the pervious subsection.

We construct the non-factorizable toroidal $\mathbb{Z}_2$~orbifold from an even dimensional torus $T^{2\ell}$ with $\ell \ge 3$, which we realize in terms of the $2\ell$~dimensional lattice $\Lambda_{2\ell}$ with lattice generators $s_A$, $A=1,\ldots,2\ell$, namely
\begin{equation} \label{eq:TorusNF}
  T^{2\ell} \simeq \mathbb{R}^{2\ell}/\Lambda_{2\ell} \ , \quad
  \Lambda_{2\ell} = \left\langle\!\left\langle \, s_1, \ldots, s_{2\ell} \, \right\rangle\!\right\rangle \ .
\end{equation}
Furthermore, we consider the involution $\iota_{\mathbb{Z}_2}$ of the torus $T^{2\ell}$, which exchanges the first $\ell$ generators with the second $\ell$ generators of the lattice $\Lambda_{2\ell}$. Explicitly, the involution $\iota_{\mathbb{Z}_2}$ is given by the lattice automorphisms 
\begin{equation} \label{eq:inv_Z2}
  \iota_{\mathbb{Z}_2}: \ s_A \mapsto 
  \begin{cases} s_{A+\ell} & \text{for $A\le\ell$ \ ,} \\ s_{A-\ell} & \text{for $A>\ell$ \ .} \end{cases}
\end{equation}  
In order for the involution  $\iota_{\mathbb{Z}_2}$ to realize a $\mathbb{Z}_2$~symmetry on the associated toroidal conformal field theory, we require that the flat toroidal target space metric $G$ and the background $B$-field $B$ are invariant with respect to this geometric $\mathbb{Z}_2$ action, i.e.,
\begin{equation}
   \iota^*_{\mathbb{Z}_2}G = G \ , \qquad \iota^*_{\mathbb{Z}_2}B = B \ . 
\end{equation}
Identifying the lattice generators $s_A$, $A=1,...,2\ell$, with a basis of tangent vectors of $T^{2\ell}$, the $\mathbb{Z}_2$~invariance of the toroidal metric $G$ implies for its symmetric components $G_{AB} = G(s_A, s_B)=G_{BA}$, $A,B=1,\ldots,2\ell$, the relations
\begin{equation} \label{eq:GZ2sym}
  G_{ab} = G_{a+\ell, b+\ell} \ , \quad G_{a+\ell, b} = G_{a, b+\ell} \quad \text{for} \ a,b=1, \ldots, \ell \ .
\end{equation}  
This means that the metric $G$ takes the form
\begin{equation}
    G = \begin{pmatrix}
        \mathfrak G & \widetilde{\mathfrak G}\\
         \widetilde{\mathfrak G}& \mathfrak G
    \end{pmatrix}\ ,\ 
\end{equation}
where $\mathfrak G,\ \widetilde{\mathfrak{G}}$ are symmetric $\ell\times \ell$ matrices.
For the anti-symmetric components $B_{AB} = B(s_A, s_B)= - B_{BA}$, $A,B=1,\ldots,2\ell$, of the $B$-field we have similar relations
\begin{equation} \label{eq:BZ2sym}
  B_{ab} = B_{a+\ell, b+\ell} \ , \quad B_{a+\ell, b} = B_{a, b+\ell} \quad \text{for} \ a,b=1, \ldots, \ell \ .
\end{equation}  
This means that the $B$-field takes the form
\begin{equation}
    B = \begin{pmatrix}
        \mathfrak B & \widetilde{\mathfrak B}\\
         \widetilde{\mathfrak B}& \mathfrak B
    \end{pmatrix}\ , \ 
\end{equation}
where $\mathfrak B,\wt{\mathfrak{B}}$ are $\ell\times\ell$ skew-symmetric matrices.
In the following, we often refer to the constructed torus $T^{2\ell}$ with the metric ~\eqref{eq:GZ2sym} and the $B$-field  \eqref{eq:BZ2sym} as the non-factorizable torus $T^{2\ell}_\text{non-fac}$, and we denote the $\mathbb{Z}_2$~orbifold associated to the involution~$\iota_{\mathbb{Z}_2}$ of the non-factorizable torus by $T^{2\ell}_\text{non-fac}/\mathbb{Z}_2$.

For ${\wt{\mathfrak{G}}}=0$, ${\wt{\mathfrak{B}}}=0$ the torus $T^{2\ell}$ factorizes into $T^\ell \times T^\ell$, where each factor  comes with the same metric~$\mathfrak{G}$ and the same $B$-field~$\mathfrak{B}$. As the $\mathbb{Z}_2$~orbifold exchanges the two tori, the non-factorizable toroidal $\mathbb{Z}_2$ conformal field theory simplifies to the $S_2$ symmetric orbifold conformal field theory arising from the product of two tori, as studied in ref.~\cite{Kames-King:2023fpa}.

Before we calculate the partition function of the conformal field theory with the orbifold target~$T^{2\ell}_\text{non-fac}/\mathbb{Z}_2$, we construct a $2^\ell$-fold cover~$\widetilde{T}^{2\ell}$ of the non-factorizable torus $T^{2\ell}_\text{non-fac}$, which is again a factorizable torus $\widetilde{T}^{2\ell} \simeq \widetilde{T}^\ell \times \widetilde{T}^\ell$ in the sense that the metric $\widetilde{G}$ and the $B$-field $\widetilde{B}$ lifted from the torus $T^{2\ell}$ becomes block diagonal. The relevant covering torus~$\widetilde{T}^{2\ell} \simeq \mathbb{R}^{2\ell}/\widetilde{\Lambda}_{2\ell}$ is described in terms of the sublattice $\widetilde{\Lambda}_{2\ell} \subset \Lambda_{2\ell}$ of index~$2^\ell$ given by 
\begin{equation}
\begin{aligned}
   \widetilde{\Lambda}_{2\ell} &=  
   \left\langle\!\left\langle \, e_1, \ldots, e_\ell, f_1, \ldots,  f_{\ell} \, \right\rangle\!\right\rangle \ , \\
    e_a &= s_a - s_{a+\ell} \ , \quad f_a = s_a + s_{a+\ell} \ ,  \quad a=1,\ldots,\ell \ .
\end{aligned}    
\end{equation}
In terms of these generators the metric $\widetilde{G}$ and the $B$-field $\widetilde{B}$ of the torus $\widetilde{T}^{2\ell}$ become block diagonal because $0=\widetilde{G}(e_a,f_b)=\widetilde{G}(f_a,e_b)$ for all $a,b=1,\ldots,\ell$. Furthermore, the metric blocks $g$ and $\tilde g$ of the two respective factors $\widetilde{T}^\ell \times \widetilde{T}^\ell$ read
\begin{equation}
  \tilde g_{ab} = \widetilde{G}(e_a,e_b) = 2 (G_{ab} -  G_{a,b+\ell}) \ ,  \quad
  {g}_{ab} = \widetilde{G}(f_a,f_b) = 2 (G_{ab} + G_{a,b+\ell}) \ .
\end{equation}
Analogously, we find for the $B$-field $\widetilde{B}$ that $0=\widetilde{B}(e_a,f_b)=\widetilde{B}(f_a,e_b)$ and that the block-diagonal entries become
\begin{equation}
  \tilde b_{ab} = \widetilde{B}(e_a,e_b) = 2 (B_{ab} - B_{a,b+\ell}) \ ,  \quad
  {b}_{ab} = \widetilde{B}(f_a,f_b) = 2 (B_{ab} + B_{a,b+\ell}) \ .
\end{equation}
In terms of the metric and $B$-field $\ell\times \ell$-blocks $g$, $\tilde g$, $b$ and $\tilde b$, we readily express the metric $G$ and $B$ of the non-factorizable torus $T^{2\ell}$ as
 \begin{equation}\label{eq:latticeGB}
	G =\frac14 \begin{pmatrix}
		g+\tilde g & g-\tilde g\\
		g-\tilde g & g+\tilde g
	\end{pmatrix} \ , \qquad
	B =\frac14 \begin{pmatrix}
 		b+\tilde b & b - \tilde b\\
 		b-\tilde b & b+\tilde b
 	\end{pmatrix} \ ,
\end{equation}
where the exhibited block structure arises in terms of the basis~\eqref{eq:TorusNF}. 
Upon inserting the specific form \eqref{eq:latticeGB} of the metric~$G$ and the $B$-field~$B$ into the matrix~$H(G,B)$, a straightforward but somewhat tedious calculation reveals that the partition function~$Z_{T^{2\ell}}(\tau; G, B)$ can be rewritten as
\begin{equation} \label{eq:Ztor2lalt} 
   Z_{T^{2\ell}_\text{non-fac}}(\tau; G, B) = 
   \frac{1}{\abs{\eta(\tau)}^{4\ell}}\sum_{\Delta \in \{0,1\}^{2\ell}} 
   \Theta_{h}(0,\tfrac12\Delta,2\tau)\ \Theta_{{\wt h}}(0,\tfrac12\Delta,2\tau) \ .
\end{equation}
Here the Siegel--Narain theta functions are defined with respect to the $\ell\times\ell$ positive definite matrices
\begin{equation} \label{eq:defhth}
  h\equiv H(\tfrac g2,\tfrac b2) \ , \qquad {\wt h}\equiv H(\tfrac {\wt g}2,\tfrac{\wt b}2) \ ,
\end{equation}
that are determined via the matrix relation~\eqref{eq:Hrel} in terms of the (rescaled) $\ell \times \ell$ matrices $g, {\wt g}, b, {\wt b}$. The modular parameter of these Siegel--Narain theta functions appearing in eq.~\eqref{eq:Ztor2lalt} is $2\tau$ as opposed to $\tau$ in eq.~\eqref{eq:Ztor2l}. As a result the modular invariance of this expression is not immediately manifest. Note that the partition function~$Z_{T^{2\ell}_\text{non-fac}}(\tau; G, B)$ as given in eq.~\eqref{eq:Ztor2lalt} is a finite sum of products of two Siegel--Narain theta functions, which only depend on $g$, $b$, and ${\wt g}$, ${\wt b}$, respectively. This is a consequence of the fact that the partition function $Z_{T^{2\ell}_\text{non-fac}}$ can alternatively be obtained from a shift orbifold of the $2^\ell$-fold covering torus ${\wt T}^{\ell} \times {\wt T}^{\ell}$, which is factorizable. 

Now we have all the ingredients at hand to spell out the partition function\linebreak $Z_{T^{2\ell}_\text{non-fac}/\mathbb{Z}_2}$ of the conformal field theory of the non-factorizable torus orbifolded with respect to the involution~\eqref{eq:inv_Z2}.
The contribution of the untwisted sector $Z^{(+)}_{T^{2\ell}_\text{non-fac}/\mathbb{Z}_2}$ to the partition function of the $\mathbb{Z}_2$ orbifold conformal field theory is obtained by projecting onto the $\mathbb{Z}_2$ invariant states on the Hilbert space, namely
\begin{equation}
    Z^{(+)}_{T^{2\ell}_\text{non-fac}/\mathbb{Z}_2} 
    = \frac12 \left( \operatorname{tr}_{\mathcal{H}_+} e^{2\pi i \tau \hat H} +  \operatorname{tr}_{\mathcal{H}_+} g\, e^{2\pi i \tau \hat H}\right) 
    = \frac12 \left( Z^{(++)}_{T^{2\ell}_\text{non-fac}/\mathbb{Z}_2} + Z^{(+-)}_{T^{2\ell}_\text{non-fac}/\mathbb{Z}_2}\right) \ ,
\end{equation}
where $\hat H$ is the Hamiltonian of the conformal field theory, $\mathcal{H}_+$ the Hilbert space of the untwisted states, and $g$ the generator of the $\mathbb{Z}_2$~orbifold action. The two summands in this expression are explicitly calculated to be 
\begin{equation} \label{eq:Ztw}
  Z^{(++)}_{T^{2\ell}_\text{non-fac}/\mathbb{Z}_2} = Z_{T^{2\ell}_\text{non-fac}}(\tau; G, B) \ , \qquad
  Z^{(+-)}_{T^{2\ell}_\text{non-fac}/\mathbb{Z}_2} = Z_{T^{\ell}}(2\tau; \tfrac{g}2, \tfrac{b}2) \ .
\end{equation}
The twisted sector $Z^{(-)}_{T^{2\ell}_\text{non-fac}/\mathbb{Z}_2}$ can be determined similarly by projecting onto the $\mathbb{Z}_2$ invariant states in the twisted sector of the $\mathbb{Z}_2$~orbifold theory, i.e.,
\begin{equation}
    Z^{(-)}_{T^{2\ell}_\text{non-fac}/\mathbb{Z}_2} 
    = \frac12 \left( \operatorname{tr}_{\mathcal{H}_-} e^{2\pi i \tau \hat H} +  \operatorname{tr}_{\mathcal{H}_-} g\, e^{2\pi i \tau \hat H}\right) 
    = \frac12 \left( Z^{(-+)}_{T^{2\ell}_\text{non-fac}/\mathbb{Z}_2} + Z^{(--)}_{T^{2\ell}_\text{non-fac}/\mathbb{Z}_2}\right) \ .
\end{equation}
Here $\mathcal{H}_-$ refers to the Hilbert space of the twisted states. However, since the whole partition function  $Z_{T^{2\ell}_\text{non-fac}/\mathbb{Z}_2}$ of the $\mathbb{Z}_2$~orbifold conformal field theory must be modular invariant, we can directly reconstruct the contributions of the twisted sector via modular transformations as follows. The contribution $Z^{(++)}_{T^{2\ell}_\text{non-fac}/\mathbb{Z}_2}$ is modular invariant by itself, because it is the partition function of the conformal field theory that is not orbifolded. The piece $Z^{(+-)}_{T^{2\ell}_\text{non-fac}/\mathbb{Z}_2}$ is not modular invariant by itself. Instead an $S$-transformation maps the contribution $Z^{(+-)}_{T^{2\ell}_\text{non-fac}/\mathbb{Z}_2}$ to $Z^{(-+)}_{T^{2\ell}_\text{non-fac}/\mathbb{Z}_2}$ and vice versa, because the $S$-transformation converts the insertion of the generator $g$ in the trace over the Hilbert space $\mathcal{H}_+$ into a twisted boundary condition without the insertion of $g$ in the trace over the Hilbert space $\mathcal{H}_-$. Thus from eq.~\eqref{eq:Ztw} we arrive at
\begin{equation}
  Z^{(-+)}_{T^{2\ell}_\text{non-fac}/\mathbb{Z}_2} = Z_{T^{\ell}}(\tfrac\tau2; \tfrac{g}2, \tfrac{b}2) \ .
\end{equation}
The last contribution is obtained from $Z^{(-+)}_{T^{2\ell}_\text{non-fac}/\mathbb{Z}_2}$ by acting with a $T$-transformation, and we find
\begin{equation}
  Z^{(--)}_{T^{2\ell}_\text{non-fac}/\mathbb{Z}_2} = Z_{T^{\ell}}(\tfrac{\tau+1}2; \tfrac{g}2, \tfrac{b}2) \ .
\end{equation}
Altogether, the combination $Z^{(+-)}_{T^{2\ell}_\text{non-fac}/\mathbb{Z}_2}+Z^{(-+)}_{T^{2\ell}_\text{non-fac}/\mathbb{Z}_2}+Z^{(--)}_{T^{2\ell}_\text{non-fac}/\mathbb{Z}_2}$ is indeed modular invariant. 

In summary, collecting all the computed individual pieces we find for the whole partition function the expression
\begin{multline} \label{eq:ZnonfacZ2}
  Z_{T^{2\ell}_\text{non-fac}/\mathbb{Z}_2}
  = \frac12 \frac{1}{\abs{\eta(\tau)}^{4\ell}}\sum_{\Delta \in \{0,1\}^{2\ell}} 
   \Theta_{h}(0,\tfrac12\Delta,2\tau)\ \Theta_{{\wt h}}(0,\tfrac12\Delta,2\tau) \\
   + \frac12 \big(Z_{T^{\ell}}(2\tau; \tfrac{g}2, \tfrac{b}2) 
   + Z_{T^{\ell}}(\tfrac{\tau}2; \tfrac{g}2, \tfrac{b}2) 
   + Z_{T^{\ell}}(\tfrac{\tau+1}2; \tfrac{g}2, \tfrac{b}2) \big) \ ,
\end{multline}
where both the first and the second line are modular invariant contributions by themselves. Note that, due to the insertion of the generator of the $\mathbb Z_2$ orbifold group, the contribution $Z^{(+-)}_{T^{2\ell}_\text{non-fac}/\mathbb{Z}_2}$ to the partition function depends only on the moduli $g,b$, which are the moduli of the fixed-point locus of the involution $\iota_{\mathbb Z_2}$.
 \subsection{Ensembles of Non-Factorizable Toroidal \texorpdfstring{$\boldsymbol{\mathbb{Z}_2}$}{Z2}~Orbifold CFTs}
 \label{sec:EnNonFacTor}
 To determine the ensemble average of the non-factorizable toroidal $\mathbb{Z}_2$~orbifold partition function $Z_{T^{2\ell}_\text{non-fac}/\mathbb{Z}_2}$, we first discuss the structure of its moduli space $\mathcal{M}_{T^{2\ell}_\text{non-fac}/\mathbb{Z}_2}$. The metric on the moduli space $\mathcal{M}_{T^{2\ell}_\text{non-fac}/\mathbb{Z}_2}$ is the Zamolodchikov metric restricted to the non-factorizable tori $T^{2\ell}$ that are invariant with respect to the action of the involution~$\iota_{\mathbb{Z}_2}$.
 
 The Zamolodchikov metric of the torus $T^{2\ell}$ is given by~\eqref{eq: Zam metric def}, which becomes in terms of the matrix \eqref{eq:Hrel}
 \begin{align} \label{eq:metTorus}
   \dd s^2_{T^{2\ell}} &= 
   \operatorname{Tr}\left( G^{-1}\dd G \,G^{-1} \dd G - G^{-1} \dd B \, G^{-1} \dd B \right) 
   \\ &= \frac12 \operatorname{Tr}\left( H^{-1} \dd H \, H^{-1} \dd H \right) \nonumber =-\frac12 \operatorname{Tr}\lr{\mathrm{d}H\dd H^{-1}}\ .
 \end{align}
Here the positive definite $2\ell \times 2\ell$ matrix $H$ is given in terms of the metric $G$ and the $B$-field $B$ according to eq.~\eqref{eq:Hrel}. Restricting the Zamolodchikov metric to the non-factorizable tori $T^{2\ell}_{\text{non-fac}}$, we insert eq.~\eqref{eq:latticeGB} and get
 \begin{equation} \label{eq:measurefactors}
 \begin{aligned}
    \dd s^2_{T^{2\ell}_\text{non-fac}} \!\!\!&=
    \operatorname{Tr}\big( g^{-1} \dd g\,g^{-1} \dd g + g^{-1} \dd b\,g^{-1} \dd b \big) 
    + \operatorname{Tr}\big( {\wt g}^{-1} \dd {\wt g}\,{\wt g}^{-1} \dd {\wt g} + {\wt g}^{-1} \dd {\wt b}\,{\wt g}^{-1} \dd {\wt b} \big) \\
    &=  \frac12 \operatorname{Tr}\big( h^{-1} \dd h\,h^{-1} \dd h \big) 
    +\frac12 \operatorname{Tr}\big( \, {\wt h}^{-1} \dd {\wt h}\,{\wt h}^{-1} \dd {\wt h}\,\big) \ .
\end{aligned}   
\end{equation}
Thus the metric factorizes locally over the moduli space of non-factorizable tori~$T^{2\ell}_{\text{non-fac}}$ into the two positive definite parts $h$ and ${\wt h}$.

Let us now analyze the global structure of the moduli space of non-factorizable tori~$T^{2\ell}_{\text{non-fac}}$. To set the stage, we first describe the moduli space $\mathcal{M}_{T^{N}}$ of conformal field theories for generic target space tori~$T^{N}$ with the positive definite $2N\times 2N$-matrices~$H$ obeying eq.~\eqref{eq:Hrelation} \cite{Siegel,siegel1955lectures}. Consider the $2N$-dimensional lattice $\Gamma$ with the even self-dual pairing $2 \Omega$ given in eq.~\eqref{eq:defOmega}. The symmetric matrix~$\Omega$ is a non-degenerate bilinear form of signature~$(N,N)$ on the $2N$-dimensional real vector space~$V=\Gamma \otimes_{\mathbb{Z}} \mathbb{R}$. Let $W_+$ be a $N$-dimensional subvector space of $V$, such that the restriction $\Omega|_{W_+}$ is positive definite. Note that the choice of $W_+$ is not unique and $W_+$ is called a majorant of $\Omega$. Furthermore, let $W_-$ be the $N$-dimensional orthogonal complement $W_- = \left\{ x\in V\,\middle|\,\Omega(x,W_+)=0 \right\}$. Then the vector space~$V$ decomposes into the direct sum
\begin{equation} \label{eq:direct}
V = W_+ \oplus W_- \ .
\end{equation}
Due to the signature of $\Omega$ the restriction $\Omega|_{W_-}$ is negative definite. From this decomposition we obtain on $V$ the positive definite symmetric bilinear form 
\begin{equation} \label{eq:HdeffromW}
    H(u,v) := \Omega(u_+,v_+) - \Omega(u_-,v_-) \ ,
\end{equation}    
with $u=u_+ + u_-$, $v=v_+ + v_-$, where $u_+, v_+ \in W_+$ and $u_-, v_- \in W_-$. Note that $H$ obeys the relation~\eqref{eq:Hrelation}, which is equivalent to $\Omega^{-1} H - H^{-1} \Omega=0$ and to
\begin{equation} \label{eq:Hrel2}
  (\Omega^{-1} + H^{-1}) (\Omega - H) = 0 \ .
\end{equation}  
Conversely, given a positive symmetric matrix $H$ obeying this matrix relation, the kernel of the second factor $\Omega - H$ defines the subvector space $W_+$ and hence the decomposition~\eqref{eq:direct} associated to the positive definite symmetric pairing $H$. 

The symmetric form $\Omega$ with signature $(N,N)$ is invariant with respect to the indefinite orthogonal group $O(N,N,\mathbb{R})$ acting on the vector space $V$, namely $\Lambda^T \Omega \Lambda = \Omega$ for any $\Lambda \in O(N,N,\mathbb{R})$. However,  the transformation on the vector space $V \mapsto \Lambda \cdot V$ acts non-trivially on the decomposition \eqref{eq:direct}, and hence on the space of positive symmetric bilinear from $H$. Conversely, Witt's theorem ensures that the group $O(N,N,\mathbb{R})$ acts transitively on the space of positive definite symmetric $2\ell \times 2\ell$ bilinear forms $H$ obeying eq.~\eqref{eq:Hrel2}. The stabilizer subgroup preserving the decomposition~\eqref{eq:direct} is $O(N,\mathbb{R}) \times O(N,\mathbb{R})$. Therefore, we find altogether that the moduli space of majorants $\mathcal{M}_\text{Maj}^{(N)}$ of $\Omega$ reads
\begin{equation}
  \mathcal{M}_\text{Maj}^{(N)} \simeq \frac{O(N,N,\mathbb{R})}{O(N,\mathbb{R}) \times O(N,\mathbb{R})} \ .
\end{equation}
As the moduli space of majorants yields a choice of metric $G$ and $B$-field $B$ according to eq.~\eqref{eq:Hrel}, it also parametrizes toroidal conformal field theories with target space~$T^{N}$. However, two majorants that are related by a lattice automorphism $O(N,N, \mathbb{Z})$ of $\Gamma$ yield equivalent toroidal conformal field theories. Therefore, we arrive at the well-known result, see, e.g., ref.~\cite{Blumenhagen:2013fgp}, that the moduli space $\mathcal{M}_{T^{N}}$ is given by
\begin{equation} \label{eq:modTorus}
    \mathcal{M}_{T^{N}} \simeq \frac{ \mathcal{M}_\text{Maj}^{(N)} }{ O(N,N,\mathbb{Z}) } 
     \simeq \left.\raisebox{-0.75ex}{$O(N,N,\mathbb{Z})$} \middle\backslash {\raisebox{0.75ex}{$O(N,N,\mathbb{R})$}} \middle\slash 
     \raisebox{-0.75ex}{$O(N,\mathbb{R}) \times O(N,\mathbb{R})$}\right. \ .
\end{equation}

Now we are ready to discuss the global structure of the moduli space $\mathcal{M}_{T^{2\ell}_\text{non-fac}}$ of conformal field theories arising from non-factorizable target space tori $T^{2\ell}_\text{non-fac}$. We parametrize the moduli space $\mathcal{M}_{T^{2\ell}_\text{non-fac}}$ in terms of majorants that admit the $\mathbb{Z}_2$~orbifold action. The involution $\iota_{\mathbb{Z}_2}$ acting on the $2\ell$-dimensional lattice $\Lambda_{2\ell}$ induces a $\mathbb{Z}_2$-action $\tilde\iota_{\mathbb{Z}_2}$ on the $4\ell$-dimensional lattice $\Gamma \simeq \Lambda_{2\ell} \oplus \Lambda_{2\ell}^*$. By construction, the involution~$\tilde\iota_{\mathbb{Z}_2}$ leaves the non-degenerate bilinear form $\Omega$ of signature $(2\ell,2\ell)$ invariant, i.e., $\tilde\iota_{\mathbb{Z}_2}^*\Omega = \Omega$. Furthermore, the vector space $V= \Gamma \otimes_{\mathbb{Z}} \mathbb{R}$ decomposes as $V= V^{(+)} \oplus V^{(-)}$, where $V^{(\pm)}$ are the $\pm 1$ eigenspaces with respect to the involution $\tilde\iota_{\mathbb{Z}_2}$. It is straightforward to check that the non-degenerate bilinear form $\Omega$ of signature $(2\ell,2\ell)$ restricts on $V^{(\pm)}$ to two non-degenerate bilinear forms $\Omega|_{V^{(\pm)}}$ both of signature~$(\ell,\ell)$. As a result the majorants compatible with the involution $\tilde\iota_{\mathbb{Z}_2}$ split as
\begin{equation} \label{eq:Vsplit_inv}
  V = W_+^{(+)} \oplus W_+^{(-)} \oplus W_-^{(+)} \oplus W_-^{(-)} \ ,
\end{equation}
where $W_\pm^{(+)} \oplus W_\pm^{(-)} = W_\pm$ and $W_+^{(\pm)} \oplus W_-^{(\pm)} = V^{(\pm)}$. The moduli spaces $\mathcal{M}_{\text{Maj},\mathbb{Z}_2}^{(2\ell)}$ of such $\mathbb{Z}_2$-equivariant majorants is parametrized by transformations that not only preserves the bilinear form $\Omega$ but also its two restrictions $\Omega|_{V^{(\pm)}}$ individually modulo those transformations that preserve the direct sum decomposition~\eqref{eq:Vsplit_inv}. Therefore, we arrive at
\begin{equation}
    \mathcal{M}_{\text{Maj},\mathbb{Z}_2}^{(2\ell)} \simeq \frac{O(\ell,\ell,\mathbb{R})\times O(\ell,\ell,\mathbb{R})}{O(\ell,\mathbb{R}) \times O(\ell,\mathbb{R})\times O(\ell,\mathbb{R}) \times O(\ell,\mathbb{R})} 
 \simeq   \mathcal{M}_{\text{Maj}}^{(\ell)} \times \mathcal{M}_{\text{Maj}}^{(\ell)}  \ .
\end{equation}

As before, in order to describe the moduli space of conformal field theories of non-factorizable target space tori~$T^{2\ell}_\text{non-fac}$, we need to further divide by those lattice automorphisms of $\Gamma$ that are compatible with the involution~$\tilde\iota_{\mathbb{Z}_2}$. These are realized by the discrete group $O(\ell,\ell,\mathbb{Z}) \times O(\ell,\ell,\mathbb{Z})$. Thus, the moduli space of the conformal field theories with non-factorizable tori $T^{2\ell}_\text{non-fac}$ as target spaces becomes
\begin{equation} \label{eq:ProdModnonfac}
\begin{aligned}
    \mathcal{M}_{T^{2\ell}_\text{non-fac}} 
    &\simeq \frac{ \mathcal{M}_{\text{Maj},\mathbb{Z}_2}^{(2\ell)} }{ O(\ell,\ell,\mathbb{Z}) \times O(\ell,\ell,\mathbb{Z}) } 
    &\simeq \left.\raisebox{-0.75ex}{$O(\ell,\ell,\mathbb{Z})^{\times 2}$} 
     \middle\backslash {\raisebox{0.75ex}{$O(\ell,\ell,\mathbb{R})^{\times 2}$}} \middle\slash 
     \raisebox{-0.75ex}{$O(\ell,\mathbb{R})^{\times 4}$}\right.  \\ 
     &\simeq \mathcal{M}_{T^{\ell}} \times \mathcal{M}_{T^{\ell}} \ .
\end{aligned}
\end{equation}
The two factors~$ \mathcal{M}_{T^{\ell}}$ of this moduli space are parameterized by the positive definite bilinear forms $h$ and $\wt h$ explicitly given in eq.~\eqref{eq:defhth}. Note that the arguments of the bilinear forms $h$ and $\wt h$ in terms of the metric $g$ and the $B$-field~$b$ and the metric $\wt g$ and the $B$-field~$\wt b$ are rescaled by a factor $\frac12$, which reflects the fact that these $\ell\times \ell$ blocks parametrize the diagonal tori $\wt T^\ell \times \wt T^\ell$ of the non-factorizable torus $T^{2\ell}_\text{non-fac}$ corresponding to the sublattice $\wt \Lambda_{2\ell}$ of index $2^\ell$.

Let us point out that the presented construction of the moduli space $\mathcal{M}_{T^{2\ell}_\text{non-fac}}$ does not cover all possible conformal field theories that can be constructed from $\mathbb{Z}_2$~orbifolds associated to the involution~$\iota_{\mathbb{Z}_2}$ acting on non-factorizable tori $T^{2\ell}_\text{non-fact}/\mathbb{Z}_2$ as defined in eq.~\eqref{eq:inv_Z2}. On top of the $B$-field~$B$ entering the majorant $H$ in eq.~\eqref{eq:Hrel}, which obeys the relations~\eqref{eq:BZ2sym}, there are additional discrete choices for the $B$-field that are invariant with respect to a $\mathbb{Z}_2$ symmetry once the discrete transformations~$O(2\ell,2\ell,\mathbb{Z})$ are taken into account. In our treatment, we only consider $B$-field configurations that are invariant under the involution $\iota_{\mathbb Z_2}$ without taking into account such discrete transformations.

The family of $S_2$-symmetric orbifold conformal field theory studied in ref.~\cite{Kames-King:2023fpa} corresponds to the points in moduli space~$\mathcal{M}_{T^{2\ell}_\text{non-fac}}$, where the equivalence classes of the majorants $h$ and $\wt h$ are equal. That is to say, the moduli space $\mathcal{M}_{T^\ell\times T^\ell/S_2}$ of the $S_2$-symmetric orbifold conformal field theory is the diagonal submoduli space
\begin{equation} \label{eq:ModS2}
   \mathcal{M}_{T^\ell\times T^\ell/S_2}
   \simeq \left\{ \, (\mathfrak{m},{\wt {\mathfrak{m}}}) \in \mathcal{M}_{T^{2\ell}_\text{non-fac}}
   \, \middle| \, \mathfrak{m} = {\wt {\mathfrak{m}}} \, \right\} \ ,
\end{equation}
where $\mathfrak{m}$ and $\wt{\mathfrak{m}}$ are equivalence classes of majorants $h$ and $\wt h$.

The metric~\eqref{eq:metTorus} of the moduli space $\mathcal{M}_{T^N}$ of a toroidal conformal field theory yields the measure $\dd \mathfrak{m}_H$, which upon integrating over the moduli space~\eqref{eq:modTorus} we normalize as
\begin{equation}
  \int_{\mathcal{M}_{T^N}} \!\!\! \dd \mathfrak{m}_H = 1 \ .
\end{equation}  
To calculate the ensemble averages of the partition functions derived in the previous section, it is necessary to average the Siegel--Narain theta functions $\Theta_H(0, \frac12 \Delta, \tau)$ defined in eq.~\eqref{eq:SNtheta} over the moduli space $\mathcal{M}_{T^N}$ as
\begin{equation}
    \left\langle \Theta_H(0,\tfrac12 \Delta, x) \right\rangle = \int_{\mathcal{M}_{T^N}} \!\!\! \dd \mathfrak{m}_H\,\Theta_H(0,\tfrac12 \Delta, x) \ .
\end{equation}    
This computation is detailed in refs.~\cite{siegel1955lectures,Dong:2021}. Here we quote the result for $N\ge 3$ and $\Delta \in \{0,1\}^{2N}$
\begin{equation} \label{eq:ThetaAverage}
  \left\langle \Theta_H(0,\tfrac12 \Delta, x) \right\rangle = 
  \begin{cases}
     \displaystyle{\frac12\sum\limits_{\genfrac{}{}{0pt}{2}{c,d\in\mathbb{Z}}{(c,d)=1}}} \frac1{|c\,x + d|^N} & \quad\text{for } 0= \Delta \in \{0,1\}^{2N}  \ , \\[4ex]
\displaystyle{\frac12\sum\limits_{\genfrac{}{}{0pt}{2}{c\in\mathbb{Z},d\in2\mathbb{Z}}{(c,d)=1}}} 
        \frac{(-1)^{\frac{d}2 \Delta^T \Omega \Delta}}{|c\,x + d|^N} & \quad\text{for }  0 \ne \Delta \in\{0,1\}^{2N}  \ .
   \end{cases}  
\end{equation}
For the average of these Siegel--Narain theta functions there are three distinct cases. The tuple~$\Delta$ is either zero or non-zero. In the latter case we distinguish between $\Delta^T \Omega \Delta$ being even or odd. Therefore, we define
\begin{equation} \label{eq:defThetaZeroPM}
\begin{aligned}
      \left\langle \Theta_H^{(0)}(x) \right\rangle &:=  \left\langle \Theta_H(0, 0, x) \right\rangle \ , \\
      \left\langle \Theta_H^{(+)}(x) \right\rangle &:=  \left\langle \Theta_H(0,\tfrac12 \Delta, x) \right\rangle 
      \quad \text{for $\Delta^T \Omega \Delta$ even, $\Delta\ne 0$} \ , \\
       \left\langle \Theta_H^{(-)}(x) \right\rangle &:=  \left\langle \Theta_H(0,\tfrac12 \Delta, x) \right\rangle 
      \quad \text{for $\Delta^T \Omega \Delta$ odd} \ .
\end{aligned}      
\end{equation}
Using eq.~\eqref{eq:ThetaAverage} and the definition for the real analytic Eisenstein series \eqref{eq:defEisenstein}, we find for the average
\begin{equation}
    \left\langle \Theta_H^{(0)}(x) \right\rangle = \frac{E_{\nicefrac{N}2}(x)}{\operatorname{Im}(\tau)^{\frac{N}2}} \ .
\end{equation}
Inserting the identities~\eqref{eq:Appendix even odd} into eq.~\eqref{eq:ThetaAverage} we arrive for the remaining averages at 
\begin{equation} \label{eq:ThetaEisensteinRels}
\begin{aligned}
  \left\langle \Theta_H^{(+)}(x) \right\rangle 
  &=\frac{1}{2^N-1}\lr{\frac{E_{\nicefrac{N}{2}}(\tfrac{x}2)}{\operatorname{Im}(\frac{x}2)^{\frac{N}{2}}}-\frac{E_{\nicefrac{N}{2}}(x)}{\operatorname{Im}(x)^{\frac{N}{2}}}}\ , \\
  \left\langle \Theta_H^{(-)}(x) \right\rangle 
 &=
  \frac2{2^{N}(2^N-1)} \frac{E_{\nicefrac{N}2}(\tfrac{x}4)}{\operatorname{Im}(\tfrac{x}4)^{\frac{N}2} } 
 -\frac{2^{N}+2}{2^{N}(2^N-1)} \frac{E_{\nicefrac{N}2}(\tfrac{x}2)}{\operatorname{Im}(\tfrac{x}2)^{\frac{N}2} } 
  + \frac1{2^N-1} \frac{E_{\nicefrac{N}2}(x)}{\operatorname{Im}(x)^{\frac{N}2} }  \ .
\end{aligned}     
\end{equation}

With these averaged Siegel--Narain theta functions at hand, we can now determine the ensemble average of the partition function~$Z_{T^{2\ell}_\text{non-fac}/\mathbb{Z}_2}$ of the $\mathbb{Z}_2$~orbifold toroidal conformal field theories based on non-factorizable tori $T^{2\ell}_\text{non-fac}$. We recall that the measure~\eqref{eq:measurefactors} and the moduli space of this ensemble of conformal field theories factorizes, which becomes manifest once we parametrize the moduli space by the majorants $h$ and $\wt h$ of eq.~\eqref{eq:defhth}. Moreover, the partition function~$Z_{T^{2\ell}_\text{non-fac}/\mathbb{Z}_2}(\tau;h,{\wt h})$ of eq.~\eqref{eq:ZnonfacZ2} is a sum of products of terms, whose factors are Siegel--Narain theta functions that depend on the two respective majorants $h$ and $\wt h$. Thus, the ensemble average factors over these sums of products as well, and we obtain
\begin{equation}
\begin{aligned}
  \left\langle \strut Z_{T^{2\ell}_\text{non-fac}/\mathbb{Z}_2}(\tau) \right\rangle 
  &= \int_{\mathcal{M}_{T^\ell}} \dd\mathfrak{m}_h \int_{\mathcal{M}_{T^\ell}} \dd\mathfrak{m}_{\wt h} 
     \ Z_{T^{2\ell}_\text{non-fac}/\mathbb{Z}_2}(\tau; h,{\wt h}) \\ 
  &= \frac12 \frac{1}{\abs{\eta(\tau)}^{4\ell}}\sum_{\Delta \in \{0,1\}^{2\ell}} 
   \left\langle\Theta_{h}(0,\tfrac12\Delta,2\tau)\right\rangle\ \left\langle\Theta_{{\wt h}}(0,\tfrac12\Delta,2\tau) \right\rangle \\
   &\quad\qquad + \frac12 \Big( \left\langle Z_{T^{\ell}}(2\tau; h) \right\rangle
   +  \left\langle Z_{T^{\ell}}(\tfrac{\tau}2; h)  \right\rangle
   +  \left\langle Z_{T^{\ell}}(\tfrac{\tau+1}2; h) \right\rangle \Big) \ .
\end{aligned}   
\end{equation}
The sum over $\Delta\in \{0,1\}^{2\ell}$ splits into the contribution $\Delta=0$, $(2^\ell-1)(2^{\ell-1}+1)$ summands with $\Delta\ne 0$ and $\Delta^T\Omega\Delta$ even, and $2^{\ell-1}(2^\ell-1)$ summands with $\Delta^T\Omega\Delta$ odd. Thus, inserting the definitions~\eqref{eq:defThetaZeroPM} and carrying out the sum over $\Delta\in \{0,1\}^{2\ell}$ we arrive at 
\begin{multline}
\left\langle \strut Z_{T^{2\ell}_\text{non-fac}/\mathbb{Z}_2}(\tau) \right\rangle 
   = \frac12 \frac{1}{\abs{\eta(\tau)}^{4\ell}} \left(  
      \left\langle \Theta_H^{(0)}(2\tau) \right\rangle^2 \right. \\
       \left.+ (2^\ell-1)(2^{\ell-1}+1) \left\langle \Theta_H^{(+)}(2\tau) \right\rangle^2 
   + 2^{\ell-1}(2^\ell-1) \left\langle \Theta_H^{(-)}(2\tau) \right\rangle^2 
      \right) \\
+ \frac12 \Big( \left\langle Z_{T^{\ell}}(2\tau; h) \right\rangle
   +  \left\langle Z_{T^{\ell}}(\tfrac{\tau}2; h)  \right\rangle
   +  \left\langle Z_{T^{\ell}}(\tfrac{\tau+1}2; h) \right\rangle \Big) \ .
\end{multline}

Finally, we express the determined average in terms of real analytic Eisenstein series using eqs.~\eqref{eq:ThetaEisensteinRels}. In order to bring the final result into a manifest modular invariant form, we apply the identify
\begin{equation}
     E_s(\tfrac{x+1}{2})=\frac{1+2^{2s-1}}{2^{s-1}}E_s(x)-E_s(2x)-E_s(\tfrac{x}{2})\ . 
\end{equation}
which is derived from the Fourier decomposition of the real analytic Eisenstein series in terms of the Hecke eigenmodes with respect to the Hecke operator $T_2$, for details, c.f., Appendix~\ref{app: eisenstein and averages}. Putting everything together, we arrive at our main result of this subsection, which is the manifest modular invariant ensemble average 
\begin{align} \label{eq:nonfact_avTN}
 & \left\langle \strut Z_{T^{2\ell}_\text{non-fac}/\mathbb{Z}_2}(\tau) \right\rangle \nonumber\\
 &\quad=\frac{1}{2\abs{\eta(\tau)}^{4\ell}\operatorname{Im}(\tau)^\ell
 \lr{2^{\ell}-1}}\bigg( \left(E_{\nicefrac{\ell}{2}}\lr{2\tau}^2+E_{\nicefrac{\ell}{2}}\lr{\tfrac{\tau}{2}}^2+E_{\nicefrac{\ell}{2}}\lr{\tfrac{\tau+1}{2}}^2\right) \nonumber\\
 &\qquad\qquad\qquad\qquad\qquad\qquad\qquad\qquad\qquad\qquad-\lr{1+2^{1-{\ell}}}E_{\nicefrac{\ell}{2}}\lr{\tau}^2\bigg)\nonumber \\
 &\quad+ \frac12\lr{\frac{E_{\nicefrac{\ell}{2}}(2\tau)}{\text{Im}(2\tau)^{\frac{\ell}{2}}\abs{\eta(2\tau)}^{2\ell}}+\frac{E_{\nicefrac{\ell}{2}}\lr{\tfrac{\tau}{2}}}{\text{Im}(\tfrac{\tau}{2})^{\frac{\ell}{2}}\abs{\eta\lr{\tfrac{\tau}{2}}}^{2\ell}}+\frac{E_{\nicefrac{\ell}{2}}\lr{\tfrac{\tau+1}{2}}}{\text{Im}(\tfrac{\tau+1}{2})^{\frac{\ell}{2}}\abs{\eta\lr{\tfrac{\tau+1}{2}}}^{2\ell}}} \ .
\end{align}
Here the expressions inside the brackets in the first and third lines are modular invariant as a consequence of the Lemma~\ref{lemma modular invariance} in Appendix~\ref{app: eisenstein and averages}. Using formulas in Appendix~\ref{sec:ModFunc} and eq.~\eqref{eq:AvZTd}, we can also write the averaged partition function as
\begin{align}\label{eq:average non fact}
  &\left\langle \strut Z_{T^{2\ell}_\text{non-fac}/\mathbb{Z}_2}(\tau) \right\rangle \nonumber\\
 &=\frac{1}{2^{\ell+1}\lr{2^\ell-1}}\left(\abs{\frac{\theta_2(\tau)}{\eta(\tau)}}^{2\ell}\braket{Z_{T^{\ell}}(2\tau)}^2+\abs{\frac{\theta_4(\tau)}{\eta(\tau)}}^{2\ell}\braket{Z_{T^{\ell}}(\tfrac{\tau}{2})}^2+\abs{\frac{\theta_3(\tau)}{\eta(\tau)}}^{2\ell}\braket{Z_{T^{\ell}}(\tfrac{\tau+1}{2})}^2\right)\nonumber \\
 &\qquad-\frac{1+2^{1-\ell}}{2\lr{2^\ell-1}}\braket{Z_{T^{\ell}}(\tau)}^2 + \frac12\lr{\braket{Z_{T^{\ell}}(2\tau)}^2+\braket{Z_{T^{\ell}}(\tfrac{\tau}{2})}^2+\braket{Z_{T^{\ell}}(\tfrac{\tau+1}{2})}^2} \ .
\end{align}
We observe that this expression for the ensemble average is consistent with the lower-dimensional regularized ensemble average stated in eq.~\eqref{eq:nonfact_av}.

Finally, notice that the ensemble average over the submoduli space $\mathcal{M}_{T^\ell\times T^\ell/S_2}$ defined in eq.~\eqref{eq:ModS2} becomes \begin{equation}
		\begin{aligned}
			\left\langle \strut Z_{T^{2\ell}_\text{non-fac}/\mathbb{Z}_2}(\tau) \right\rangle 
			&= \frac12 \frac{1}{\abs{\eta(\tau)}^{4\ell}}
			\left\langle\Theta_{h}(0,0,\tau)\Theta_{{ h}}(0,0,\tau) \right\rangle \\
			&\quad\qquad + \frac12 \Big( \left\langle Z_{T^{\ell}}(2\tau; h) \right\rangle
			+  \left\langle Z_{T^{\ell}}(\tfrac{\tau}2; h)  \right\rangle
			+  \left\langle Z_{T^{\ell}}(\tfrac{\tau+1}2; h) \right\rangle \Big) \ ,
		\end{aligned}   
	\end{equation}
which is the ensemble average of the partition function for the product of two equal tori $T^\ell$ orbifolded by the permutation $S_2$ as calculated in ref.~\cite{Kames-King:2023fpa}.
\section{Dual Holographic Chern--Simons Theory} \label{sec:Bulk} 
In this section we discuss the ensemble averages of conformal field theories of the type calculated in the previous section from a three-dimensional holographic dual bulk perspective. In the section~\ref{sec:Bulk_prod} we analyze ensemble averages of two-dimensional conformal field theories that arise from products of families of conformal field theories. For such products the ensemble average factorizes, which raises a conundrum about their dual holographic interpretation. We discuss possible scenarios for such cases. In  section~\ref{section: ensemble averages of toroidal z2 cft} we examine the calculated ensemble averages of the previous section, which exhibit similar phenomena as the ensemble averages of products of families of two-dimensional conformal field theories.

\subsection{Products of Conformal Field Theories}
\label{sec:Bulk_prod}
Consider two families of two-dimensional unitary conformal field theories $\CFT{a}$, $a=1,2$, which are parametrized by (local) coordinates $\mathfrak{m}_a$ of their moduli spaces~$\mathcal{M}_a$. Their partition functions are defined in the usual manner
\begin{equation}
   Z_a(\tau;\mathfrak{m}_a) = \operatorname{Tr}_{\mathcal{H}_a} \left( q^{L_0(\mathfrak{m}_a)} {\bar q}^{\bar L_0(\mathfrak{m}_a)} \right) \ , 
   \quad a=1,2 \ ,
\end{equation}   
where $\mathcal{H}_a$ are the Hilbert spaces of states, $L_0$ and $\bar L_0$ are the moduli-dependent holomorphic and anti-holomorphic degree zero Virasoro generators of the conformal field theories $\CFT{a}$, and $q=e^{2\pi i \tau}$. 

We are assuming that both moduli spaces $\mathcal{M}_a$, $a=1,2$, have fixed finite dimension $d_a = \dim \mathcal{M}_a$. This implies that for generic values of $\mathfrak{m}_a$ the conformal field theories $\CFT{a}$ possess $d_a$ marginal operators, i.e., $d_a$ primary fields of conformal dimension $(h,\bar h)=(1,1)$.\footnote{The conformal field theory $\CFT{a}$ may have additional primary fields of conformal dimension $(h,\bar h) = (1,1)$ for some values of $\mathfrak{m}_a$. These primaries are either exactly marginal operators for non-generic values of $\mathfrak{m}_a$ or they are not exactly marginal. In the former case it means that a new stratum of families of conformal field theories is attached to $\mathcal{M}_a$ at this non-generic point in moduli space. In the latter case deformations with respect to such operators are obstructed at higher orders.} Furthermore, suppose that the moduli spaces $\mathcal{M}_a$ have finite volumes
\begin{equation}
   \operatorname{Vol}(\mathcal{M}_a) = \int_{\mathcal{M}_a} \dd \mu(\mathfrak{m}_a) \ , \quad a=1,2 \ ,
\end{equation}   
with respect to the measure $\dd \mu(\mathfrak{m}_a)$ induced from the Zamolodchikov metrics of the conformal field theories $\CFT{a}$. Then we can define the ensemble averages
\begin{equation}
  \left\langle Z_a(\tau) \right\rangle = \frac1{\operatorname{Vol}(\mathcal{M}_a)}
  \int_{\mathcal{M}_a} \dd \mu(\mathfrak{m}_a) \, Z_a(\tau;\mathfrak{m}_a)  \ ,
\end{equation}
which we assume to be also finite.

Finally, assume that both ensemble averages $\left\langle Z_a(\tau) \right\rangle$, $a=1,2$, enjoy a holographic interpretation in terms of a three-dimensional bulk action $S_a[g,\phi_a]$ on three-dimensional spaces $M_3$ with a single toroidal boundary component $\partial M_3=T^2_\tau$ with complex structure~$\tau$. Due to the conformal symmetry at the boundary $T^2_\tau$, the holographic duality implies further that the metric approaches three-dimensional (Euclidean) anti-de Sitter space --- that is to say a hyperbolic three-space --- at the asymptotic region of the boundary $\partial M_3=T^2_\tau$. The boundary conditions of the remaining fields $\phi_a|_{T^2_\tau}$ conform with the spectrum of the ensemble average of conformal field theories. The described holographic interpretation implies at the quantitative level that the partition functions of the three-dimensional bulk theory calculate the ensemble average of the families of conformal field theories, i.e.,
\begin{equation}
    \left\langle Z_a(\tau) \right\rangle = \int \mathcal{D}[g,\phi_a] \, e^{-S_a[g,\phi_a]}\ . 
\end{equation}
Here the RHS is a functional integral over bulk field configurations with specified boundary conditions, bulk metrics that on the boundary give rise to $T^2_\tau$, which possibly includes sums over topologies. For example, in refs.~\cite{Maloney:2020nni,Afkhami-Jeddi:2020ezh}, the path integral over the metric~$g$ localizes to a sum over hyperbolic geometries $M_3$ with asymptotic boundary conditions~$\partial M_3 = T^2_\tau$ with complex structure modulus $\tau$. In the context of the ensemble averages of higher genus partition functions of symmetric product orbifolds considered in ref.~\cite{Kames-King:2023fpa} smooth bulk geometries that are not handlebodies become relevant. See also refs.~\cite{Yin:2007at, Yin:2007gv}.
Now we consider the product of the two two-dimensional conformal field theories $\CFT{1\otimes 2} \equiv \CFT{1} \times \CFT{2}$, which by construction again yields a family of unitary conformal field theories parametrized by $(\mathfrak{m}_1,\mathfrak{m}_2)\in\mathcal{M}_1 \times \mathcal{M}_2$. Since the Zamolodchikov metric and hence the moduli space measure also factorize, the ensemble average of the family of product conformal field theories is just the product of the averages of its factors, i.e.,
\begin{equation} \label{eq:PartProduct}
   \left\langle Z_{1\otimes 2}(\tau) \right\rangle = \left\langle Z_{1}(\tau) \right\rangle \, \left\langle Z_{2}(\tau) \right\rangle \ .
\end{equation}
From the dual holographic perspective the observed factorization of the partition function poses a puzzle, as the partition function of a possible holographic dual three-dimensional bulk theory is also required to factorize. Assuming that a holographic description exists at all, we discuss in the following scenarios for possible bulk interpretations of such ensemble averages:
\begin{itemize}
\item Since the ensemble average $\left\langle Z_{1\otimes 2}(\tau) \right\rangle$ factorizes, a possible interpretation for a three-dimensional dual description is obtained in terms of the three-dimensional action $S_{1\otimes 2} = S_1[g_1,\phi_1] + S_2[g_2,\phi_2]$. In this setup the two metrics $g_1$ and $g_2$ are distinct on the two three-spaces $M_3^{(1)}$ and $M_3^{(2)}$, and both three-spaces $M_3^{(a)}$, $a=1,2$, should have a common asymptotic toroidal boundary $T^2_\tau=\partial M_3^{(1)}=\partial M_3^{(2)}$ --- as depicted in Fig.~\ref{fig: commnon bdy bulks} --- on which the two metrics $g_1$ and $g_2$ coincide asymptotically. Then by construction the holographic correspondence becomes
\begin{equation} \label{eq:Prod3dBulk}
\begin{aligned}
   \left\langle Z_{1\otimes 2}(\tau) \right\rangle
   &= \int \mathcal{D}[g_1,g_2,\phi_1,\phi_2] \, e^{-S_1[g_1,\phi_1]-S_2[g_2,\phi_2]} \\
   &= \int \mathcal{D}[g_1,\phi_1] \, e^{-S_1[g_1,\phi_1]} \int \mathcal{D}[g_2,\phi_2]\,e^{-S_2[g_2,\phi_2]} \\
   &= \left\langle Z_{1}(\tau) \right\rangle \, \left\langle Z_{2}(\tau) \right\rangle \ .
\end{aligned}   
\end{equation}
Thus, the holographic dual arises from two distinct three-dimensional bulk theories that are glued together at a common asymptotic boundary. 
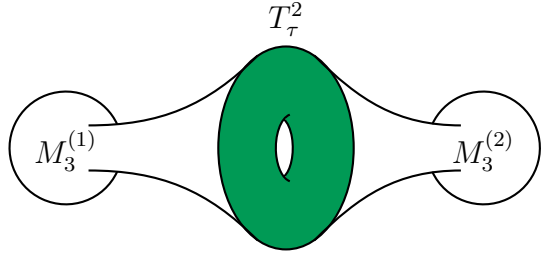
\begin{figure}
    \centering
   	\begin{tikzpicture}[scale = 1.5]
	\begin{scope}[xshift = 5.25cm, scale = 0.3]
		\fill[ForestGreen, opacity = 1] (0,0) [partial ellipse = 0:360:2 and 3];
		\fill[white] (0.2,0) [partial ellipse = 115:245:0.5 and 1];
		\fill[white] (-0.2,0) [partial ellipse = -70:70:0.4 and 0.9];
		\draw[thick] (0,0) [partial ellipse = 0:360:2 and 3];
		\draw[thick] (0.2,0) [partial ellipse = 100:260:0.5 and 1];
		\draw[thick] (-0.2,0) [partial ellipse = -70:70:0.4 and 0.9];
		\node[above] at (0,3) {$T^2_\tau$};
	\end{scope}
\begin{scope}[scale = 1.00,xshift=5.5cm]
	
	\draw[thick] (0,0.82) to[out =-45, in = 180] (1.3,0.2);
	\draw[thick] (0,-0.82) to[out = 45, in = 180] (1.3,-0.2);
	
	\node[below] at (1.5,0.25) {$ M_3^{(2)}$};
	
	\draw[thick] (-0.5,0.82) to[out =-135, in = 0] (-2.,0.2);
	\draw[thick] (-0.5,-0.82) to[out =135, in = 0] (-2.,-0.2);
	
	\node[below] at (-2.2,0.25) {$ M_3^{(1)}$};
	
	\draw[thick] (1.5,0) [partial ellipse = -155:155:0.5 and 0.5];

	\draw[thick] (-2.2,0) [partial ellipse = 335:25:0.5 and 0.5];
	
\end{scope}
\end{tikzpicture}
    \caption{Bulk manifolds $M_3^{(1)},M_3^{(2)} $ sharing the same boundary.}
    \label{fig: commnon bdy bulks}
\end{figure}
\item The product conformal field theory $\CFT{1\otimes 2}$ could be a $(d_1+d_2)$-dimensional subspace $\mathcal{M}_1 \times \mathcal{M}_2$ of a higher-dimensional moduli space $\mathcal{M}_\text{total}$, i.e, $\mathcal{M}_1 \times \mathcal{M}_2 \subset \mathcal{M}_\text{total}$ with $\dim \mathcal{M}_\text{total}> d_1 + d_2$. Such a scenario occurs, if the conformal field theory~$\CFT{1\otimes 2}$ has in addition to the exactly marginal operators $\varphi^{(1)}_i\otimes\mathbf{1}$, $i=1,\ldots,d_1$, and $\mathbf{1}\otimes \varphi^{(2)}_j$, $j=1,\ldots,d_2$, further exactly marginal operators. The operators~$\mathbf{1}$ are the identity operators, and $\varphi^{(1)}_i$ and $\varphi^{(2)}_j$ are the marginal operators of conformal weight $(h,\bar h)=(1,1)$ of the respective conformal field theories $\CFT1$ and $\CFT2$. The additional marginal operators are of the form $\psi^{(1)} \otimes \psi^{(2)}$, where the operators $\psi^{(1)}$ and $\psi^{(2)}$ of the conformal field theories $\CFT1$ and $\CFT2$ have conformal dimensions~$(h^{(1)},\bar h^{(1)})$ and $(h^{(2)},\bar h^{(2)})$, such that $h^{(1)} + h^{(2)}=\bar h^{(1)} + \bar h^{(2)}=1$. These additional exactly marginal operators $\psi^{(1)} \otimes \psi^{(2)}$ parametrize directions in $\mathcal{M}_\text{total}$ that are normal to the product subspace $\mathcal{M}_1\times\mathcal{M}_2$.\footnote{A given two-dimensional product conformal field theory represents a point $p \in \mathcal{M}_1\times\mathcal{M}_2 \subset \mathcal{M}_\text{total}$. The set of exactly marginal operators form a basis of the tangent space $T_p\mathcal{M}_\text{total}$ at the point $p \in \mathcal{M}_\text{total}$. At any point $p\in \mathcal{M}_1\times\mathcal{M}_2$ this tangent bundle splits as
$T_p\mathcal{M}_\text{total} \simeq T_p(\mathcal{M}_1\times\mathcal{M}_2)\oplus N_p(\mathcal{M}_1\times\mathcal{M}_2)$, where the two summands denote the tangent and normal bundle of $\mathcal{M}_1\times\mathcal{M}_2$ relative to the embedding space $\mathcal{M}_\text{total}$, respectively. The exactly marginal operators that parametize deformations within the submoduli space $\mathcal{M}_1\times\mathcal{M}_2$ reside in the tangent bundle $T_p(\mathcal{M}_1\times\mathcal{M}_2)$, whereas those exactly marginal operators that parametrize deformations that wander off into the bigger moduli space $\mathcal{M}_\text{total}$ have non-vanishing components in the normal bundle $N_p(\mathcal{M}_1\times\mathcal{M}_2)$.}

Generically, it is not expected that the larger moduli space $\mathcal{M}_\text{total}$ exhibits a product structure (unless enforced by symmetry). Therefore, the ensemble average of the partition function $Z_\text{total}(\tau; \mathfrak{m}_\text{total})$ of the total family of conformal field theories $\CFT{\text{total}}$ does not exhibit the product structure any longer
\begin{equation}
  \left\langle Z_\text{total}(\tau)\right\rangle = \frac1{\operatorname{Vol}(\mathcal{M}_\text{total})}
  \int_{\mathcal{M}_\text{total}}\!\!\! \dd \mu(\mathfrak{m}_\text{total})\, Z_\text{total}(\tau; \mathfrak{m}_\text{total}) \ ,
\end{equation}
where we assume that the volume of $\mathcal{M}_\text{total}$ is finite and that the integral over the partition function~$\mathcal{M}_\text{total}$ converges.

Since the dimension of the moduli space $\mathcal{M}_\text{total}$ is higher than the subspace $\mathcal{M}_1\times\mathcal{M}_2$, the contribution to the ensemble average arising from this subspace has measure zero. Its contribution is therefore not relevant for the total ensemble average. As a result a holographic dual formulation for the conformal field theory $\CFT{\text{total}}$ does not need to reflect a product structure any longer, and the average $\left\langle Z_\text{total}(\tau)\right\rangle$ can possibly arise from a conventional three-dimensional bulk theory on a three-dimensional space $M_3$ with an asymptotic boundary component~$T^2_\tau$.
\item Finally, let us remark that even if the product of conformal field theories does not give rise to additional exactly marginal operators, the products of conformal field theories could be part of a larger ensemble, in which the product moduli space $\mathcal{M}_1\times\mathcal{M}_2$ arises as a connected component. If the connected component has measure zero in this larger moduli space $\mathcal{M}_\text{total}$, then the holographic dual description --- if it exists at all --- should be given the entire moduli space $\mathcal{M}_\text{total}$, which does not need to reflect the product structure~\eqref{eq:PartProduct} unless the entire moduli space $\mathcal{M}_\text{total}$ is again a product itself. 
Such a scenario is conceivable, if for instance the ensemble $\mathcal{M}_\text{total}$ of conformal field theories consists of different strata of different dimensions. Furthermore, if the product stratum $\mathcal{M}_1\times \mathcal{M}_2$ is of lower dimension than any of the other strata, it does not contribute to ensemble averages over the entire moduli space. It would be interesting to find explicit examples of this type. 
\end{itemize}

\subsection{Ensemble Averages of Toroidal \texorpdfstring{$\boldsymbol{\mathbb{Z}_2}$}{Z2}~Orbifold CFTs}\label{section: ensemble averages of toroidal z2 cft}
Let us illustrate the general considerations in the previous subsection with a few examples of families of toroidal conformal field theories and their $\mathbb{Z}_2$ orbifolds that are discussed in this work:
\paragraph{Ensemble Average of Products of Toroidal CFTs:}
As proposed in refs.~\cite{Maloney:2020nni, Afkhami-Jeddi:2020ezh} the holographic dual description of the ensemble average~\eqref{eq:AvZTd} of the family of two-dimensional conformal field theories $\CFT{T^d}$ with a toroidal target space~$T^d$ is given in terms of a $U(1)^{d}\times U(1)^{d}$ Chern--Simons theory on three-dimensional hyperbolic handlebodies with a single toroidal asymptotic boundary component~$T^2_\tau$. The bulk gauge one-forms $A_a$, ${\wt A}_a$, $a=1,\ldots,\ell$, of the gauge group $U(1)^d\times U(1)^d$ obey the holomorphic and anti-holomorphic boundary conditions \cite{Benjamin:2021wzr}
\begin{equation}
  A_{a}|_{T^2_\tau} = \partial_z X_a(z,\bar z) \, \dd z \ , \qquad  {\wt A}_{a}|_{T^2_\tau} = \partial_{\bar z} X_a(z,\bar z) \, \dd \bar{z} \ ,
  \qquad a=1,\ldots,d \ .
\end{equation}
Here $z$ are holomorphic coordinates at the boundary~$T^2_\tau$ and $X_a(z,\bar z)$, $a=1,\ldots,d$, are the free bosonic fields of the two-dimensional conformal field theory on $T^2_\tau$, which parametrize the target space torus $T^d$. Note that the fields $\partial_z X_a$ and $\partial_{\bar z}X_a$ are primary fields of conformal dimension $(h,\bar h)=(1,0)$ and $(h,\bar h)=(0,1)$, respectively. Moreover, there are $d^2$ primaries $\partial_z X_a \partial_{\bar z}X_b$, $a,b=1,\ldots,d$, of conformal dimension $(h,\bar h) = (1,1)$ that make up for the exactly marginal operators of the $d^2$-dimensional moduli space $\mathcal{M}_{T^d}$.

Now we consider the product family of conformal field theory $\CFT{T^\ell\otimes T^m}$ of the two families of toroidal conformal field theories $\CFT{T^\ell}$ and $\CFT{T^m}$ with target tori~$T^\ell$ and $T^m$. The ensemble average over the product moduli space $\mathcal{M}_{T^\ell} \times \mathcal{M}_{T^{m\phantom{\ell}}}$ yields\footnote{Products of real analytic Eisenstein series from products of Narain conformal field theories or from subloci associated to such Narain products are also discussed in  ref.~\cite{Benjamin:2021ygh}.} 
\begin{equation}
  \left\langle Z_{T^\ell\otimes T^m}(\tau) \right\rangle = 
  \frac{ E_{\ell/2}(\tau) E_{m/2}(\tau)}{\text{Im}(\tau)^{\frac{\ell+m}2}\abs{\eta(\tau)}^{2(\ell+m)}} =
  \left\langle Z_{T^{\ell}}(\tau) \right\rangle \left\langle Z_{T^{m\phantom{\ell}}}(\tau) \right\rangle \ .
\end{equation}  
We associate the exactly marginal operators $\partial_z X_a \partial_{\bar z}X_b \otimes \mathbf{1}$, $a,b=1,\ldots,\ell$ and $\mathbf{1} \otimes \partial_z {\wt X}_a \partial_{\bar z}{\wt X}_b$, $a,b=1,\ldots,m$, to the $(\ell^2 + m^2)$-dimensional moduli space $\mathcal{M}_{T^\ell} \times \mathcal{M}_{T^{m\phantom{\ell}}}$ of the product family. Here $X_a$, $a=1,\ldots,\ell$, and ${\wt X_a}$, $a=1,\ldots,m$, are the free bosons of the two toroidal conformal field theories. 

The product family of conformal field theories $\CFT{T^\ell\otimes T^m}$ has $2\ell m$ additional exactly marginal operators, namely
\begin{equation}
  \partial_z X_a  \otimes \partial_{\bar z}{\wt X}_b  \ , \qquad
  \partial_{\bar z} X_a  \otimes \partial_{z}{\wt X}_b\ , \quad a=1,\ldots,\ell \ , \ \ b=1,\ldots,m \ .
\end{equation}  
These additional exactly marginal operators extend the moduli space $\mathcal{M}_{T^\ell} \times \mathcal{M}_{T^{m\phantom{\ell}}}$ to the larger moduli space $\mathcal{M}_{T^D}$ with $D=\ell + m$ of dimension $D^2 = \ell^2 + m^2 + 2\ell m$. Thus, the product family of conformal field theories $\CFT{T^\ell\otimes T^m}$ with moduli space $\mathcal{M}_{T^\ell}\times\mathcal{M}_{T^{m\phantom{\ell}}}$ naturally embeds into the larger family of toroidal conformal field theories with target space torus $T^D$ and moduli space $\mathcal{M}_{T^D}$, which yields the ensemble average~\eqref{eq:AvZTd} with the discussed holographic dual in terms of the bulk $U(1)^D \times U(1)^D$ Chern--Simons gauge theory. 

\paragraph{Ensemble Average of Factorizable Toroidal $\boldsymbol{\mathbb{Z}_2}$ Orbifold CFTs:}
The ensemble average of the partition function of the family of factorizable toroidal $\mathbb{Z}_2$~orbifold conformal field theories studied in subsection~\ref{sec:FacTorOrbTD} factorizes as established in eq.~\eqref{eq:EAfacTD} because the moduli space $\mathcal{M}_{T^{D}/\mathbb{Z}_2}$ factorizes as $\mathcal{M}_{T^\ell} \times \mathcal{M}_{T^m}$, c.f., eq.~\eqref{eq:ProdModfac}.

As opposed to the moduli space of the product conformal field theory $\CFT{T^\ell \otimes T^m}$ discussed in the previous paragraph, for generic points $(\mathfrak{m}_1,\mathfrak{m}_2)\in \mathcal{M}_{T^\ell} \times \mathcal{M}_{T^{m\phantom{\ell}}}$ the family of conformal field theories $\CFT{T^\ell/\mathbb{Z}_2 \otimes T^m}$ does not have any primary fields with conformal dimension $(h,\bar h)=(1,1)$ that are of the form $\psi^{(1)} \otimes \psi^{(2)}$, where $\psi^{(1)}$ and $\psi^{(2)}$ are primaries of the respective conformal field theory factors. Therefore, the moduli space $\mathcal{M}_{T^\ell} \times \mathcal{M}_{T^m}$ is not a lower-dimensional slice of a higher-dimensional embedding moduli space. 

Hence, the moduli space of the product conformal field theory $\CFT{T^\ell/\mathbb{Z}_2 \otimes T^m}$ does not naturally extend to a larger family of conformal field theories. However, the factorized ensemble average $\left\langle Z_{T^{\ell}/\mathbb{Z}_2 \otimes T^{m}}(\tau) \right\rangle=\left\langle Z_{T^{\ell}/\mathbb{Z}_2}(\tau) \right\rangle\left\langle Z_{T^{m}}(\tau) \right\rangle$ of the partition function can still be obtained from the holographic duals of the conformal field theory factors $\CFT{T^\ell/\mathbb{Z}_2}$ and $\CFT{T^m}$ along the lines of eq.~\eqref{eq:Prod3dBulk}. The three-dimensional holographic duals of these two factors of families of conformal field theories are proposed in refs.~\cite{Maloney:2020nni,Afkhami-Jeddi:2020ezh} and in ref.~\cite{Benjamin:2021wzr}, respectively.
\paragraph{Ensemble Average of Non-Factorizable Toroidal $\boldsymbol{\mathbb{Z}_2}$ Orbifold CFTs:}
The moduli space $\mathcal{M}_{T^{2\ell}/\mathbb{Z}_2}$ for the family of conformal field theories resulting from non-factorizable toroidal $\mathbb{Z}_2$~orbifold conformal field theories studied in subsection~\ref{sec:EnNonFacTor} again factorizes, c.f., eq.~\eqref{eq:ProdModnonfac}.  However, the partition function~$Z_{T^{2\ell}_\text{non-fac}}$ does not factorize, and hence the ensemble average is neither of the factorized form. As a consequence, the boundary conditions of the fields of a prospective holographic dual three-dimensional bulk theory does not decompose into two sectors that respect the product form of the moduli space $\mathcal{M}_{T^{2\ell}/\mathbb{Z}_2}$. 
Nevertheless, the resulting ensemble average~\eqref{eq:nonfact_avTN} becomes a finite sum of products of real analytic Eisenstein series. It would be interesting to propose a holographic dual, which possibly consists of non-trivial topological sectors, similarly as the vortex sectors considered in refs.~\cite{Benjamin:2021wzr,Kames-King:2023fpa}. We hope to come back to this question in the future.

\section{Conclusions and Outlook} \label{sec:con} 
In this work, we analyze and explicitly construct $\mathbb Z_2$ orbifolds --- and their moduli spaces --- of toroidal conformal field theories arising from topologically distinct classes involutions~$\iota_{\mathbb Z_2}$ of their target space tori. This serves as a new testing ground for examples of ensemble average holography in the sense of refs.~\cite{Maloney:2020nni, Afkhami-Jeddi:2020ezh}. Using the Siegel--Weil formula for averaging Siegel--Narain theta functions, we obtain explicit formulas for the averages in terms of sums of products of real analytic Eisenstein series.

\paragraph{Factorizable Toroidal Target Spaces:}
The family of factorizable toroidal CFTs with moduli space $\mathcal{M}_{T^\ell} \times \mathcal{M}_{T^{m\vphantom{\ell}}}$ naturally embeds into the larger moduli space $\mathcal{M}_{T^{\ell + m}}$. Put differently, the factorizable family of CFTs has additional exactly marginal operators that can deform the theory away from the product $\mathcal{M}_{T^\ell}\times \mathcal{M}_{T^{m\vphantom{\ell}}}$. Averaging the partition functions only over the moduli~$\mathcal{M}_{T^\ell} \times \mathcal{M}_{T^{m\vphantom{\ell}}}$ requires a dual holographic description of the product type as well as discussed around eq.~\eqref{eq:Prod3dBulk}. However, including deformations with respect to all exactly marginal operators and considering the larger ensemble~$\mathcal{M}_{T^{\ell + m}}$ is well motivated here, as the original Narain correspondence is obtained~\cite{Maloney:2020nni, Afkhami-Jeddi:2020ezh}.

For the family of factorizable $\mathbb{Z}_2$ orbifold conformal field theories the moduli space is again identified with $\mathcal{M}_{T^\ell} \times \mathcal{M}_{T^{m\vphantom{\ell}}}$. However, at a generic point in moduli space there are no further exactly marginal operators that canonically extend the factorizable moduli space. Hence, averaging over the product moduli space is a canonical choice. 

For these two cases, we argue in section \ref{section: ensemble averages of toroidal z2 cft} that one can give a holographic interpretation to the ensemble average of the partition function by considering bulk theories that share a common boundary as illustrated in Fig. \ref{fig: commnon bdy bulks}. This notion also appears in ref.~\cite{Kames-King:2023fpa} and these arguments resemble the holomorphic factorization considerations that appear in refs.~\cite{Maloney:2007ud,Yin:2007gv}.
\paragraph{Non-Factorizable Toroidal Target Spaces:}

The main focus of this work are families of conformal field theories and their $\mathbb{Z}_2$~orbifolds with the toroidal target space that are non-factorizable. That is to say that starting with a two-dimensional torus target space~$T^2$ which is not a Cartesian product of circles, we define a $\mathbb Z_2$ action and calculate the partition function of the $\mathbb Z_2$ orbifold and the regularized ensemble average. For the low-dimensional target torus $T^2$, we find an expression~\eqref{eq: non fact Z2 partition function} for the partition function solely in terms of partition functions of free bosonic conformal field theories on the circle.

We generalize this construction to higher dimensional target space tori. In order to determine ensemble averages, we express the partition function in terms of Siegel--Narain theta functions instead of expressing them in terms of low-dimensional toroidal partition functions (as for the target space $T^2$).\footnote{We expect that in principle an interesting formula for the partition function of non-factorizable $\mathbb{Z}_2$ orbifolds in terms of partition functions of lower-dimensional tori can be derived.} To calculate averages over the moduli space $\mathcal{M}_{T^{2\ell}_\text{non-fac}}$ associated to this ensemble, we employ the Siegel--Weil formula. We derive a manifest modular-invariant expression~\eqref{eq:average non fact} for the ensemble average of the paritition function in terms of a sum of products of averages of toroidal partition functions of lower dimension. For generic members of conformal field theories in this ensemble all exactly marginal operators describe deformations of the moduli space $\mathcal{M}_{T^{2\ell}_\text{non-fac}}$, and therefore averaging over this moduli space takes into account all deformations consistent with the $\mathbb{Z}_2$~orbifold action in the considered family of conformal field theories. This moduli space has an interesting connection to the moduli space $\mathcal{M}_{\text{Sym}^N(T^\ell)}$ of family of symmetric product orbifold conformal field theories $\text{Sym}^2\lr{T^\ell}$ discussed in ref.~\cite{Kames-King:2023fpa}.\footnote{More generally, ref.~\cite{Kames-King:2023fpa} studies ensemble averages of symmetric product orbifolds conformal field theories $\text{Sym}^N(T^\ell)$ with $N\geq 2$.}Namely, $\mathcal{M}_{\text{Sym}^2\lr{T^\ell}}$ is the subslice of $\mathcal{M}_{T^{2\ell}_\text{non-fac}}$, where the tensor product of two identical toroidal CFTs is orbifolded by $S_2\simeq \mathbb Z_2$. One can consider deformations associated to exactly marginal operators that deform this product structure to the larger moduli space of non-factorizable target space tori $T^{2\ell}_\text{non-fac}$ discussed in the present work. 
	
An interesting and possibly non-trivial calculation/construction is to generalize our work for the non-factorizable target space to non-factorizable $S_N$ orbifolds, where $S_N$ denotes the permutation group of $N$ elements. This represents a generalization of the $\text{Sym}^N(T^\ell)$ orbifolds considered in ref.~\cite{Kames-King:2023fpa}.
	
We have also discussed in the main text that one can add to our construction discrete choices of $B$-field that are invariant under $\mathbb Z_2$ action only once the discrete duality transformations $O(2\ell,2\ell;\mathbb Z)$ are taken in to account. It would be nice to see how explicit one can be with averages of toroidal orbifolds that include these choices and whether these choices have a bulk meaning.
\paragraph{Holographic interpretation:}
For the ensemble of factorizable $\mathbb{Z}_2$~orbifold toroidal conformal field theories, we propose a possible holographic interpretation in terms of perviously discussed Chern--Simons theories on bulk manifolds that share a common boundary. It would be important to quantitively further check such a proposal and to discuss its implications.

Formulating a holographic dual for the non-factorizable case seems even more challenging. While the moduli space still factors for non-factorizable $\mathbb{Z}_2$ orbifold toroidal conformal field theories, we have not put forward a proposal for a possible dual bulk theory. However, the derived analytic expression for the ensemble average of the partition function in terms of products of Eisenstein series suggests that the $S_2$ permutation symmetry might play an important role in the bulk theory as well. Alternatively, we can view the non-factorizable toroidal $\mathbb{Z}_2$~orbifolds as arising from a shift orbifold (c.f., for instance ref.~\cite{Wendland:2000ye}) of the factorizable toroidal $\mathbb{Z}_2$~orbifolds. This perspective might also shed light on a possible holographic bulk interpretation in the future.
\paragraph{Other orbifolds, discrete torsion and supersymmetry:}
One can think of generalizing our construction by considering $\mathbb Z_N$, $N=3,4,\ldots$, or other discrete groups such as $\mathbb Z_2\times \mathbb Z_2$ or $\mathbb Z_N\times \mathbb Z_M$. In particular, the latter cases are interesting because they admit discrete torsion.
On the level of the partition function of orbifolds of conformal field theories, discrete torsion amounts to assigning suitable group-theoretic phase factors to its various orbifold sectors such that partition function is modular invariant \cite{Blumenhagen:2013fgp,VAFA198659,FONT1989272}. A concrete example for an orbifold conformal field theory with discrete torison is given by orbifold toroidal conformal field theories of the type \cite{FONT1989272}
\begin{equation}
    \frac{T^D\times T^D \times T^D}{\mathbb Z_2\times \mathbb Z_2} \ .
\end{equation}
It would be interesting to study such classes of orbifolds with discrete torsion from the scope of our current work.

Another future direction is to add supersymmetry to our setting along the lines of ref.~\cite{Kames-King:2023fpa}, where the original Narain averaged duality \cite{Maloney:2020nni,Afkhami-Jeddi:2020ezh} is generalized to include supersymmetry.

\bigskip
\subsection*{Acknowledgments.}
We would like to thank 
Giorgos Batzios,
Hirosi Ooguri,
and
Maik Sarve
for interesting discussions and correspondences. JKK and SF thank the Mainz Institute for Theoretical Physics (MITP) for hospitality at various stages of this project.
AK is supported
by the Onassis Foundation as an Onassis Scholar (Scholarship ID: F ZO 030/2 (2020-
2021)).
HJ, AK and IGZ are supported by the Cluster of Excellence Precision Physics, Fundamental Interactions, and Structure of Matter (PRISMA+, EXC 2118/1) within the German Excellence Strategy (Project-ID 390831469).  SF\ acknowledges support by
the Bonn Cologne Graduate School of Physics and Astronomy (BCGS). JKK is supported by the National Centre of Competence in Research SwissMAP.

\newpage
\appendix
\section{Theta Functions and Toroidal Partition Functions} \label{sec:ModFunc} 
\subsection*{Some Theta Function Transformations}
Here we collect some useful definitions and relations related to theta functions and the Dedekind eta function. Theta functions with characteristics $\alpha,\beta$ are defined as 
\begin{equation}
    \thetas{\alpha}{\beta}(z|\tau) = \sum_{n\in\mathbb{Z}}\text{exp}\lr{i\pi (n+\alpha)^2\tau+2\pi i (n+\alpha)(z+\beta)}\ .
\end{equation}
Here $\tau=\tau_1+i\tau_2\in\mathbb H$ is the modular parameter of the genus one Riemann surface that is defined in the upper half plane $\mathbb H = \{\left.x+i y\right| y>0; x,y \in \mathbb R\}$, and $z$ is a point on this Riemann surface.

We are particularly interested in the theta functions~$\thetas{\alpha}{\beta}(\tau)\equiv\thetas{\alpha}{\beta}(0|\tau) $. The following definitions appear in partition functions in the main text
\begin{equation}
    \thetas{1/2}{0}(\tau) = \theta_2 (\tau)\ ,\quad \thetas{0}{0}(\tau) = \theta_3 (\tau)\ ,\quad \thetas{0}{1/2}(\tau) = \theta_4 (\tau)\ .
\end{equation}
	Note these theta functions transform under modular transformations as:
	\begin{equation}
		\begin{aligned}
			\theta_2(\tau+1)&=e^{\frac{i\pi}{4}}\theta_2(\tau) \ ,& 
			\theta_2\lr{-\tfrac1\tau}&=\sqrt{-i\tau}\theta_4(\tau)\\
			\theta_3(\tau+1)&=\theta_4(\tau)\ ,&  \theta_3\lr{-\tfrac1\tau}&=\sqrt{-i\tau}\theta_3(\tau)\\
			\theta_4(\tau+1)&=\theta_3(\tau)\ ,& \theta_4\lr{-\tfrac1\tau}&=\sqrt{-i\tau}\theta_2(\tau)\\
			\eta(\tau+1)&=e^{\frac{i\pi}{12}}\eta(\tau)
			\ ,&
			\eta\lr{-\tfrac1\tau}&=\sqrt{-i\tau}\eta(\tau)
		\end{aligned}
	\end{equation}
The Dedekind eta function is defined as
\begin{equation}
    \eta(\tau) = q^{1/24} \prod_{n=1}^{\infty} (1-q^n)\ , \qquad q = e^{2\pi i \tau}\ .
\end{equation}
The following identities are useful:
\begin{align}\label{theta identities}
	\frac{\theta_2(\tau)}{\eta(\tau)}=\frac{2\eta(2\tau)^2}{\eta(\tau)^2}\ ,\qquad \frac{\theta_4(\tau)}{\eta(\tau)}=\frac{\eta(\tau/2)^2}{\eta(\tau)^2}\ ,\qquad \frac{\theta_3(\tau)}{\eta(\tau)}=\frac{\eta((\tau+1)/2)^2}{e^{\pi i/12}\eta(\tau)^2}
\end{align}
Starting with the first identity in eq.~\eqref{theta identities}, we can prove the others by modular transformations, namely
\begin{equation}
		\frac{\theta_2(\tau)}{\eta(\tau)}=\frac{2\eta(2\tau)^2}{\eta(\tau)^2}\substack{S\\ \longrightarrow} 	\frac{\theta_4(\tau)}{\eta(\tau)}=\frac{2\eta(-2/\tau)^2}{(-i\tau)\eta(\tau)^2}=\frac{2(-i\tau/2)\eta(\tau/2)^2}{(-i\tau)\eta(\tau)^2}=\frac{\eta(\tau/2)^2}{\eta(\tau)^2}\ ,
\end{equation}
\begin{equation}
	\frac{\theta_4(\tau)}{\eta(\tau)}=\frac{\eta(\tau/2)^2}{\eta(\tau)^2}\substack{T\\ \longrightarrow} \frac{\theta_3(\tau)}{\eta(\tau)}=\frac{\eta((\tau+1)/2)^2}{e^{\pi i/12}\eta(\tau)^2}\ .
\end{equation}
\subsection*{Toroidal Partition Functions and Theta Functions}
Consider a two-dimensional CFT on a Riemann surface of genus $g=1$ and target space a $D$-dimensional torus $T^D$. The moduli of this theory are the metric $G_{MN}$ and the anti-symmetric $B$-field $B_{MN}$. We sometimes denote these collectively as 
\begin{equation}
\mathfrak m = \{G_{MN}, B_{MN}\} \ .
\end{equation}
The partition function is
\begin{equation}
	Z_{T^D}(\tau;\mathfrak m) = \frac{1}{\abs{\eta(\tau)}^{2D }}\sum_{M,W\in \mathbb Z^D}\text{exp}\lr{F(\tau;\mathfrak m)}
\end{equation} 	
with
\begin{multline}\label{eq: F exponent, partition function}
	F(\tau;\mathfrak m) = 	2\pi i \tau_1 M_A W^A-\alpha'\pi  \tau_2 \Big(M_A G^{AB} M_B + \frac{1}{\lr{\alpha'}^2} W^A G_{AB} W^B -\\-\frac{2}{\alpha'}W^A B_A^{\ B}M_B-\frac{1}{\lr{\alpha'}^2} W^A B_A^{\ B}B_{BC}W^C\Big) \ .
\end{multline}   
  In terms of the theta functions defined in section \ref{sec: section 2 thetas} the partition function can be written as
  \begin{equation}
      Z_{T^D}(\tau;\mathfrak m) = \frac{1}{\abs{\eta(\tau)}^{2 D }}\Theta_{H\lr{G,B}}(0,0,\tau)\ . 
  \end{equation}
\subsection*{Average of Theta Functions}
For the average of Siegel--Narain theta functions~\eqref{eq:SNtheta} with rational characteristic $b \in \mathbb{Q}^{2D}$ over the moduli space $\mathcal{M}_{T^D}$ of $2D\times 2D$ matrices $H$ (c.f., the discussion in section~\ref{sec:EnNonFacTor}), one gets \cite{Dong:2021}
\begin{equation} \label{eq:thetaaverage}
    \left\langle \Theta_H(0,b, x) \right\rangle = \int_{\mathcal{M}_{T^D}} \!\!\! \dd \mathfrak{m}_H\,\Theta_H(0,b, x)
    =\sum_{(c,d)=1,c\ge 0}\frac{\gamma_{d\cdot b}}{\abs{c\tau+d}^{D}}\,e^{2\pi i d c^* \, b^\mu \Omega_{\mu\nu}b^\nu}\ ,
\end{equation}
where $(c,d)$ denotes the common greatest divisor of the integers $c$ and $d$ and the $2D\times 2D$-matrix $\Omega$ is given in eq.~\eqref{eq:defOmega}. The symbol $\gamma_v$ for any $v\in \mathbb{R}^{2D}$ is defined as
\begin{equation}
   \gamma_v = \begin{cases} 1 & \text{for } v\in \mathbb{Z}^{2D} \ , \\
   0 & \text{else} \ .
   \end{cases}
\end{equation}
The integer $c^*$ is part of a B\'ezout pair $(c^*,d^*)$ obeying $c \, c^* + d \, d^* = (c,d) = 1$, which exists for any coprime integers $c$ and $d$ by B\'ezout's Lemma. Note that the average~\eqref{eq:thetaaverage} is well-defined for any choice of B\'ezout's pair $(c^*,d^*)$. For more details on this formula, see ref.~\cite{Dong:2021}.
\newpage
\section{Details on Non-Factorizable Tori} \label{sec:NF details} 
To calculate the partition function, we need the quantity from \eqref{eq: F exponent, partition function}, $F(\tau;\mathfrak m)$ evaluated for the metric and $B$ field above (hence $\mathfrak m$ depends on $g,\wt g, b, \wt b$). To do so, first split the momentum and winding modes $M,W$ into two $D$-dimensional vectors, like so
\begin{equation}
 		M_A W^A = \lr{m_1,..,m_\ell,\widetilde m_1,...,\widetilde m_\ell }\renewcommand{\arraystretch}{0.5}\begin{pmatrix}
 			w^1\\
 			.\\
 			.\\
 			.\\
 			\widetilde w^1\\
 			.\\
 			.\\
 			.\\
 			\widetilde w^\ell
 		\end{pmatrix} = m_a w^a + \widetilde m_a \widetilde w^a 
 	\end{equation}
and define the quantities
	\begin{align}
 		\mathfrak{r}_{\pm,a} = m_a \pm  \wt m_a\ , \qquad \mathfrak l_{\pm}^a =  w^a \pm  \wt w^a\ .
 	\end{align}
  These definitions, together with the identities
  \begin{equation}
 		\begin{pmatrix}
 			m^\top,\wt{m}^\top
 		\end{pmatrix}\begin{pmatrix}
 		A + \wt{A} & A - \wt{A}\\
 			A - \wt{A} & A + \wt{A}
 		\end{pmatrix}\begin{pmatrix}
 		m\\
 		\wt{m}
 		\end{pmatrix} = \begin{pmatrix}
 		m+\wt{m}
 		\end{pmatrix}^\top A \begin{pmatrix}
 		m+\wt{m}
 		\end{pmatrix} +  \begin{pmatrix}
 		m-\wt{m}
 		\end{pmatrix}^\top \wt A \begin{pmatrix}
 		m-\wt{m}
 		\end{pmatrix}\\
 	\end{equation}
  \begin{equation}
 		\begin{pmatrix}
 			w^\top,\wt{w}^\top
 		\end{pmatrix}\begin{pmatrix}
 			A + \wt{A} & A - \wt{A}\\
 			A - \wt{A} & A + \wt{A}
 		\end{pmatrix}\begin{pmatrix}
 			m\\
 			\wt{m}
 		\end{pmatrix} = \begin{pmatrix}
 			w+\wt{w}
 		\end{pmatrix}^\top A \begin{pmatrix}
 			m+\wt{m}
 		\end{pmatrix} +  \begin{pmatrix}
 			w-\wt{w}
 		\end{pmatrix}^\top \wt A \begin{pmatrix}
 			m-\wt{m}
 		\end{pmatrix}\\
 	\end{equation}
  where $A, \wt A$ are $\ell\times \ell$ matrices, enable us to write the partition function in a nice form. Essentially these identities rely on the fact that 
  \begin{equation}\lr{P^{-1}}^\top
      \begin{pmatrix}
 			A + \wt{A} & A - \wt{A}\\
 			A - \wt{A} & A + \wt{A}
 		\end{pmatrix}P^{-1}=\begin{pmatrix}
 		    A&0\\
            0&\wt A
 		\end{pmatrix}\ , \qquad P = \begin{pmatrix}
 		    1&1\\
            1&-1
 		\end{pmatrix}\ .
  \end{equation}
  We get
  \begin{equation}
      F(\tau;\mathfrak m (g,b;\wt g, \wt b)) =  F_+(\tau;\mathfrak m(g,b))+ F_-(\tau;\mathfrak m(\wt g,\wt b))\ ,
  \end{equation}
  with 
  \begin{multline}\label{eq: Fplus}
 		F_+(\tau;\mathfrak m) = 	\pi i \tau_1 \lr{\mathfrak{r}_{+,a}\mathfrak{l}_{+}^a}-\alpha'\pi  \tau_2 \Big(\mathfrak{r}_{+,a} g^{ab} \mathfrak r_{+,b}+\frac{1}{4\lr{\alpha'}^2}\lr{\mathfrak{l}_+^a g_{ab} \mathfrak l_+^b}\\-\frac{1}{\alpha'}\Big(\mathfrak l_+^a b_a^{\ b}\mathfrak{r}_{+,b}\Big) -\frac{1}{4\lr{\alpha'}^2}\lr{\mathfrak{l}_+^a(b)^2_{ab} \mathfrak l_+^b}\Big) 
 	\end{multline}
 	and
 	\begin{multline}\label{eq: Fminus}
 		F_-\lr{\tau;\mathfrak m}= 	\pi i \tau_1 \lr{\mathfrak{r}_{-,a}\mathfrak{l}_{-}^a}-\alpha'\pi  \tau_2 \Big(\mathfrak{r}_{-,a} \wt g^{ab} \mathfrak r_{-,b}+\frac{1}{4\lr{\alpha'}^2}\lr{\mathfrak{l}_-^a \wt g_{ab} \mathfrak l_-^b}\\-\frac{1}{\alpha'}\Big(\mathfrak l_-^a\wt b_a^{\ b}\mathfrak{r}_{-,b}\Big)-\frac{1}{4\lr{\alpha'}^2}\lr{\mathfrak{l}_-^a(\wt b)^2_{ab} \mathfrak l_-^b}\Big)\ .
 	\end{multline}
  Here, the dependence of the moduli $\mathfrak m$ is on $g,b$ or $\wt g, \wt b$. The upshot is that $ F(\tau;\mathfrak m)$ splits into the sum of two quantities, one of which depends only on $g,b$ and the other only on $\wt g,\wt b$ (as far as target space moduli are concerned). Note that  $\mf r_\pm, \mf l_\pm$ are vectors in $\mathbb{Z}^\ell$ and take even or odd values in a correlated way. This means
  \begin{equation}\label{eq:sectors even odd}
 		\mf r_\pm = 2r_\pm + p\ , \qquad\mf l_\pm = 2l_\pm + q \end{equation}
are vectors in $\mathbb{Z}^\ell$ and $p,q\in \{0,1\}^\ell$. Plugging \eqref{eq:sectors even odd} into \eqref{eq: Fplus} and \eqref{eq: Fminus}, we obtain (written in matrix notation)
 \begin{multline}
 	F_+(\tau;\mathfrak m) = 2 \pi i (2\tau_1)\lr{r_++\frac p2}^\top \lr{l_++\frac q 2}-\alpha'\pi (2\tau_2) 2\Big(\lr{r_++\frac p2}^\top g^{-1}\lr{r_++\frac p2}\\ +\frac1{4\lr{\alpha'}^2} \lr{l_++\frac q 2}^\top g \lr{l_++\frac q 2}-\frac1{\alpha'} \lr{l_++\frac q 2}^\top b g^{-1}\lr{r_++\frac p2}\\-\frac1{4\lr{\alpha'}^2}\lr{l_++\frac q 2}^\top bg^{-1}b\lr{l_++\frac q 2}\Big)
 \end{multline}
 	and similarly for $F_-\lr{\tau;\mathfrak m}$. The partition function can be written as
 	\begin{equation}
 		 Z_{T^{2\ell}_\text{non-fac}}(\tau; G, B) = 
   \frac{1}{\abs{\eta(\tau)}^{4\ell}}\sum_{\Delta \in \{0,1\}^{2\ell}} 
   \Theta_{h}(0,\tfrac12\Delta,2\tau)\ \Theta_{{\wt h}}(0,\tfrac12\Delta,2\tau) \ .
 	\end{equation}
\section{Real Analytic Eisenstein Series}\label{app: eisenstein and averages}
\subsection*{Real Analytic Eisenstein Identities}
In the calculation of the ensemble average~\eqref{eq:ThetaAverage} we use for the real analytic Eisenstein series the identities
\begin{equation} \label{eq:Appendix even odd}
\begin{aligned}
  \displaystyle{\frac12\sum\limits_{\genfrac{}{}{0pt}{2}{c\in\mathbb{Z},d\in2\mathbb{Z}}{  (c,d)=1}}} 
        \frac{1}{|c\,x + d|^N}&=
  \frac{1}{2^N-1}\lr{\frac{E_{\nicefrac{N}{2}}(\tfrac{x}2)}{\operatorname{Im}(\frac{x}2)^{\frac{N}{2}}}-\frac{E_{\nicefrac{N}{2}}(x)}{\operatorname{Im}(x)^{\frac{N}{2}}}}\ , \\
 \displaystyle{\frac12\sum\limits_{\genfrac{}{}{0pt}{2}{c\in\mathbb{Z},d\in2\mathbb{Z}}{  (c,d)=1}}} 
        \frac{(-1)^{\frac{d}2 }}{|c\,x + d|^N}  &=\frac1{2^N-1}\left(
  \frac2{2^{N}} \frac{E_{\nicefrac{N}2}(\tfrac{x}4)}{\operatorname{Im}(\tfrac{x}4)^{\frac{N}2} } 
 -\frac{2^{N}+2}{2^{N}} \frac{E_{\nicefrac{N}2}(\tfrac{x}2)}{\operatorname{Im}(\tfrac{x}2)^{\frac{N}2} } 
  +  \frac{E_{\nicefrac{N}2}(x)}{\operatorname{Im}(x)^{\frac{N}2} }  \right)\ .
\end{aligned}     
\end{equation}
where in the sum the symbol $(c,d)$ denotes the greatest common divisor of $c$ and $d$, i.e, $(c,d)=1$ says that $c$ and $d$ are coprime. To show these identities we start with a useful lemma:
\begin{lemma} \label{lem:sums}
    Let $\alpha$ be any positive integer. Then for any integers $c$ and $d$ the following two conditions are equivalent: \\
    \hspace*{10ex} (i)\ $(c,2^\alpha d)=1$, \qquad\qquad
    (ii)\ $c$ odd and $(c,d)=1$.
\end{lemma}
\begin{proof}
    If $c$ and $2^\alpha d$ are coprime, then both $c$, $2^\alpha$ and $c$, $d$ are coprime. Hence, $c$ odd and $(c,d)=1$. Conversely, if $c$ is odd then $c$, $2^\alpha$ are coprime. As $c$ and $d$ are coprime as well, altogether $c$, $2^\alpha d$ must be coprime and hence $(c,2^\alpha d)=1$. 
 \end{proof}

We now show the first real analytic Eisenstein identity~\eqref{eq:Appendix even odd} explicitly, and we start with the calculation
\begin{equation} \label{eq:CalcFirstId}
\begin{aligned}
	\sum_{\substack{(c,d)=1\\d \text{ even}}}\abs{c\cdot x+d}^{-N}&=\sum_{\substack{(c,2d)=1}}\abs{c\cdot x+2d}^{-N}=2^{-N}\sum_{\substack{(c,d)=1\\c \text{ odd}}}\abs{c\cdot \tfrac{x}{2}+d}^{-N}\\
	&=2^{-N}\sum_{\substack{(c,d)=1}}\abs{c\cdot \tfrac{x}{2}+d}^{-N}-2^{-N}\sum_{\substack{(c,d)=1\\c \text{ even}}}\abs{c\cdot \tfrac{x}{2}+d}^{-N}\\
	&=2^{-N}\sum_{\substack{(c,d)=1}}\abs{c\cdot \tfrac{x}{2}+d}^{-N}-2^{-N}\sum_{\substack{(2c,d)=1}}\abs{c\cdot x+d}^{-N}\\
	&=2^{-N}\sum_{\substack{(c,d)=1}}\abs{c\cdot \tfrac{x}{2}+d}^{-N}-2^{-N}\sum_{\substack{(c,d)=1\\d \text{ odd}}}\abs{c\cdot x+d}^{-N}\\
	&=2^{-N}\sum_{\substack{(c,d)=1}}\abs{c\cdot \tfrac{x}{2}+d}^{-N}-2^{-N}\sum_{\substack{(c,d)=1}}\abs{c\cdot x+d}^{-N}\\
 &\qquad\qquad\qquad\qquad\qquad\quad\,+2^{-N}\sum_{\substack{(c,d)=1\\d \text{ even}}}\abs{c\cdot x+d}^{-N}\ ,
\end{aligned}
\end{equation}
where the summations are manipulated time and again using Lemma~\ref{lem:sums}. Solving in this expression for~$\sum_{\substack{(c,d)=1\\d \text{ even}}}\abs{c\cdot x+d}^{-N}$ yields
\begin{equation}\label{eq: result for even}
	\sum_{\substack{(c,d)=1\\d \text{ even}}}\abs{c\cdot x+d}^{-N} = \frac{1}{2^N-1}\lr{\sum_{\substack{(c,d)=1}}\abs{c\cdot \tfrac{x}{2}+d}^{-N}-\sum_{\substack{(c,d)=1}}\abs{c\cdot x+d}^{-N}}\ .
\end{equation}
Inserting the definition of the real analytic Eisenstein series~\eqref{eq:defEisenstein}, we arrive at the first identity~\eqref{eq:Appendix even odd}.

For the second identity~\eqref{eq:Appendix even odd} our derivation is similar but a bit more tedious, because we first split the sum over $d$ into positive and negative contributions. This can be achieved by introducing an auxiliary summation index $d'$ for the even integers $d$, which discriminates between the positive and negative part by setting $d=4d'$ and $d=2(2d'+1)$. After splitting the sum in this way, we perform a similar calculation as in eq.~\eqref{eq:CalcFirstId} to obtain the second identity~\eqref{eq:Appendix even odd}.
\subsection*{Hecke Operators and Modularity}
The real analytic Eisenstein series $E_s(x)$ are eigenfunctions of the Hecke operators. This means
\begin{equation}
	T_jE_{s}(x) :=\frac{1}{\sqrt{j}}\sum_{\substack{ad=j,d>0\\0\leq b\leq d-1}}E_s\lr{\tfrac{ax+b}{d}}= \frac{\sigma_{2s-1}(j)}{j^{s-\tfrac12}}E_{s}(x)\ ,
\end{equation}
 where $\sigma_n(x)=\sum_{\left.d\right|x }d^n$ is the sum of positive divisor function. We have 
\begin{align}
	T_2 E_{s}(x) &= \frac{1}{\sqrt2}\lr{E_s(2x)+E_s(\tfrac{x}{2})+E_s(\tfrac{x+1}{2})}\\
	\sigma_{2s-1}(2) &= 1+2^{2s-1}\ .
\end{align}

Let us finally also state a useful lemma, which we use in the main text to identify manifest modular invariant combinations of real analytic Eisenstein series:
\begin{lemma}\label{lemma modular invariance}
	Let $f(x)$ be a modular invariant function $f(x)$ with respect to the modular group $\operatorname{PSL}(2,\mathbb{Z})$, which acts on the argument $x$ by M\"obius transformations. Then the function $g(x)$, given by
	\begin{equation}
		g(x) = f(2x)+f(\tfrac{x}{2})+f(\tfrac{x+1}{2})\ ,
	\end{equation}
    is modular invariant.
\end{lemma}
\begin{proof}
    The modular group $\operatorname{PSL}(2,\mathbb{Z})$ is generated by the standard generators $T$ and $S$ that map $x$ to $x+1$ and $x$ to $-\frac{1}{x}$, respectively. For the generator $T$ we calculate $g(x+1) = f(2x+2) + f(\tfrac{x+1}2) + f(\tfrac{x+2}2) = g(x)$ because $f(2x+2)=f(2x)$ and $f(\tfrac{x+2}2)=f(\tfrac{x}2)$ by the modularity of $f$. For the generator $S$ we find 
    $$
      g(-\tfrac1x) = f(-\tfrac2x) + f(-\tfrac1{2x}) + f(\tfrac{x-1}{2x})
      = f(\tfrac{x}2) + f(2x) + f(-\tfrac{2x}{x-1}) \ ,  
    $$
    where for the second equal sign the modularity of the function $f$ is again used. Furthermore, by modularity of $f$, we have for the last term in this equation
    $$ 
      f(-\tfrac{2x}{x-1})= f(-\tfrac{2}{x-1}-2) = f(-\tfrac{2}{x-1}) = f(\tfrac{x-1}2) = f(\tfrac{x+1}2) \ ,
    $$ 
    which demonstrates altogether that $g(-\tfrac1x)=g(x)$. 
    Thus, $g(x)$ is invariant with respect to both generators $T$ and $S$, and hence is a modular invariant function.
\end{proof}
\newpage
\bibliographystyle{utphys.bst}
\bibliography{3dbib}
\end{document}